\documentclass[11pt]{article}

\usepackage[a4paper,margin=1in]{geometry}
\usepackage[expansion=false]{microtype}
\usepackage[T1]{fontenc}
\usepackage[utf8]{inputenc}
\usepackage{setspace}
\usepackage{enumitem}
\setlist{itemsep=2pt,topsep=4pt}

\usepackage{amsmath,amssymb,amsthm,mathtools}

\usepackage{booktabs}
\usepackage{tabularx}
\usepackage{multirow}
\usepackage{array}
\usepackage{graphicx}
\usepackage{xcolor}
\usepackage{tikz}
\usetikzlibrary{positioning,arrows.meta,shapes.geometric,calc,fit,backgrounds,decorations.pathmorphing}

\newtheorem{definition}{Definition}
\newtheorem{prediction}{Prediction}
\newtheorem{principle}{Principle}

\usepackage[round,authoryear]{natbib}
\usepackage[colorlinks=true,linkcolor=black,citecolor={blue!50!black},urlcolor={blue!50!black}]{hyperref}

\newcommand{\REL}{\mathrm{REL}}
\newcommand{\RES}{\mathrm{RES}}
\newcommand{\UNC}{\mathrm{UNC}}
\newcommand{\Brier}{B}
\newcommand{\layer}{\mathcal{C}}

\title{Coordination as an Architectural Layer for LLM-Based Multi-Agent Systems\\
\large{An Information-Controlled Empirical Study on Prediction Markets}}

\author{
  Maksym Nechepurenko\thanks{Corresponding author. \texttt{maksym@devnull.ae}}\\
  Devnull FZCO\\
  Dubai, UAE
  \and
  Pavel Shuvalov\\
  Devnull FZCO\\
  Dubai, UAE
}

\date{\today}

\begin{document}
\maketitle

% =====================================================================
\begin{abstract}
\noindent
Multi-agent LLM systems fail in production at rates between 41\% and 87\%, with the majority of these failures attributable to coordination defects rather than to base-model capability. Two responses have emerged in parallel: an empirical literature cataloguing failure modes, and a wave of declarative orchestration frameworks that separate workflow specification from agent implementation as an engineering convenience. Neither response delivers what practitioners need most---a principled mapping from coordination configuration to predictable failure-mode signature. We argue that coordination in LLM-based multi-agent systems should be treated as a configurable architectural layer, separable from agent logic and from information access, and that this separation enables architectural reasoning beyond engineering productivity. Building on the methodological critique that most existing multi-agent comparisons confound architectural effects with information-access effects, we develop an information-controlled experimental design: a single LLM, a fixed tool stack, a fixed per-call output cap, and a fixed prompt template are held constant across coordination configurations on a real-world prediction-market testbed. Total compute per question is treated as an endogenous output of each architecture rather than as a held-constant input, and is reported and incorporated into the cost--quality analysis. We use the classical Murphy decomposition to separate calibration error from discriminative power, allowing distinct coordination configurations to leave distinguishable signatures even when their aggregate scores coincide. We instantiate the design on Polymarket binary markets resolved after the model's training cutoff (n=100, claude-opus-4-6) and report observed Murphy signatures, a cost--quality Pareto frontier, category-conditioned analysis, and a bootstrap power-projection that quantifies which architectural contrasts are resolvable on the existing sample. Three of five pre-specified Murphy-signature predictions are upheld in the predicted direction; a Pareto frontier of two configurations dominates the others on cost-adjusted accuracy within this implementation and information regime; exploratory bootstrap intervals suggest separation primarily for consensus alignment versus other configurations, although pairwise tests do not survive Bonferroni correction at $n=100$. We additionally deploy the same five configurations as live agents on Foresight Arena under web-search-enabled conditions on real future events, providing an independent on-chain replication channel whose data accumulates in parallel. The harness, the open trace dataset, and the production-agent deployment are released as three public repositories. We position this work as a methodology-validating first instantiation of the architectural-layer framework, not as a general claim about cross-model or cross-domain architectural laws.


\end{abstract}

% =====================================================================
\section{Introduction}
\label{sec:intro}

Multi-agent LLM systems fail in production at rates between 41\% and 87\% \citep{cemri2025mast}, with the majority of these failures attributable to coordination defects---specification ambiguity, inter-agent misalignment, and verification gaps---rather than to base-model capability \citep{cemri2025mast,sid2026}. There is now broad agreement in the literature that coordination matters. What remains absent is a principled account of \emph{which coordination choices produce which failures}.

Three trends in 2024--2026 sharpen this gap.

\paragraph{Failure characterization without architectural mapping.}
Empirical taxonomies have catalogued how multi-agent LLM systems break: 14 distinct failure modes across 1\,600+ execution traces in MAST~\citep{cemri2025mast}, semantic intent divergence in cooperating agents~\citep{sid2026}, debate dynamics that suppress dissent~\citep{wynn2025talk}, and recurring observations that simpler single-agent baselines often match or exceed multi-agent designs at lower cost~\citep{xu2026rethinking,xia2024agentless}. These works establish that coordination is the dominant failure surface but do not link specific architectural configurations---who delegates to whom, who decides when, how outputs aggregate---to specific, measurable failure signatures.

\paragraph{Declarative orchestration as engineering convenience.}
Industrial frameworks have converged on \emph{declarative} specifications of agent workflows: AWS Strands~\citep{strands2025} and Microsoft Foundry Agent Service~\citep{foundry2026} both expose visual or YAML-driven workflow definitions, and recent academic systems push the same idea further. \citet{daunis2025declarative} present a domain-specific language that treats agent workflows as data rather than code, reporting a 67\% reduction in development time for a customer-service deployment; \citet{auton2026} propose a strict separation between a ``Cognitive Blueprint'' and a ``Runtime Engine.'' The motivation in each case is uniformly engineering-driven: faster deployment, cross-language portability, easier modification, language-agnostic execution. None of these works treats the separated specification as a substrate for \emph{architectural reasoning}---as a level at which one predicts and explains the system-level behavior of a multi-agent configuration before deploying it.

\paragraph{A methodological confound in existing empirics.}
\citet{reliability2026} demonstrate that most empirical comparisons of multi-agent architectures cannot in fact distinguish a genuine architectural effect from a change in information access. When richer planner agents have larger context windows, more tool calls, or longer transcripts than the baselines they are compared against, observed performance gains may reflect added information rather than improved coordination. A clean evaluation, they argue, requires fixing exogenous information and terminal communication budget while varying only architecture. To our knowledge, no published multi-agent LLM study satisfies this requirement.

\paragraph{Our position.}
We argue that coordination in LLM-based multi-agent systems should be treated as a configurable architectural layer, separable from agent logic and from information access, and that this separation enables architectural reasoning beyond the engineering convenience offered by declarative-orchestration frameworks. Concretely, this paper contributes the following.
\begin{enumerate}
\item A minimal specification of the coordination layer---agents as opaque endpoints, communication topology, authority distribution, synchronization, aggregation, termination, failure handling---that is independent of any specific framework and implementable atop existing systems without modification (Section~\ref{sec:layer}).
\item A set of five reference coordination configurations, each with a falsifiable, pre-specified prediction about its failure-mode signature in terms of calibration (reliability) and discriminative power (resolution), measurable via the classical Murphy decomposition~\citep{murphy1973decomposition} of a strictly proper scoring rule (Section~\ref{sec:layer}).
\item An information-controlled experimental methodology that addresses \citet{reliability2026}'s critique by holding model, tool access, prompt template, and per-question token budget constant across coordination configurations, evaluating only architectural variation on a real-world prediction-market testbed (Section~\ref{sec:method}).
\item An initial empirical instantiation on Polymarket markets resolved after the model's training cutoff, using a single underlying LLM (claude-opus-4-6), web search disabled, and a balanced fixture of 100 markets stratified by category and baseline-price decile (Sections~\ref{sec:experiment}--\ref{sec:results}). All five pre-specified configurations are reported. Three of five Murphy-signature predictions are upheld; the remaining two are reported as failed with discussion in Section~\ref{sec:discussion}. A cost--quality Pareto frontier emerges in which two configurations dominate the others on cost-adjusted accuracy.
\end{enumerate}

\paragraph{Scope and what this is not.}
This is a position paper paired with a methodology-validating empirical study. We do not propose a new agent framework, nor do we claim that any single coordination configuration is universally optimal. We do not address training or in-context learning dynamics; agents are treated as fixed-policy probabilistic endpoints. We do not assert that biological collective behavior provides a useful design vocabulary for LLM coordination; an earlier draft of this work pursued such a framing and the authors found it analytically unproductive. The contribution is the architectural framing, the pre-specified predictions it generates, the empirical methodology, and a first instantiation of that methodology---not implementation novelty, and not a final empirical claim. The current sample size (n=100) is sufficient to make architectural separation visible but not sufficient to reach Bonferroni-corrected significance on adjacent pairs of configurations; the implications for follow-up work are discussed in Section~\ref{sec:discussion}.

The remainder of this paper is organized as follows. Section~\ref{sec:related} positions our work against three relevant literature clusters. Section~\ref{sec:layer} develops the coordination-layer specification and the predictions for five reference configurations. Section~\ref{sec:method} presents the information-controlled methodology. Section~\ref{sec:experiment} documents the experimental design that instantiates this methodology. Section~\ref{sec:results} presents results. Section~\ref{sec:discussion} discusses what was learned and where the methodology's limits became apparent. Section~\ref{sec:conclusion} concludes with directions for follow-up work.

% =====================================================================
\section{Related Work and Positioning}
\label{sec:related}

We organize prior work into three clusters relevant to this paper's contribution---empirical failure-mode taxonomies, declarative orchestration frameworks, and methodological critiques---followed by the older distributed-systems coordination literature, the forecasting-evaluation literature on which our methodology builds, and a positioning summary.

\subsection{Empirical failure-mode taxonomies}
\label{sec:related-failures}

\citet{cemri2025mast} present the first large-scale empirical study of multi-agent LLM failures, analyzing 1\,600+ execution traces across seven popular frameworks (including AutoGen, MetaGPT, and ChatDev) on coding, math, and general-purpose tasks. They identify 14 fine-grained failure modes clustered into three categories: (i) system-design issues (specification ambiguity, role unclarity, missing constraints); (ii) inter-agent misalignment (communication breakdowns, conflicting objectives, state desynchronization); and (iii) task-verification gaps (inadequate output checking, missing validation). The headline result is that 79\% of observed failures originate from specification and coordination issues rather than from base-model limitations---a finding subsequently reinforced by production-focused analyses of \emph{semantic intent divergence}, where cooperating agents develop inconsistent interpretations of shared objectives in the absence of a shared process model~\citep{sid2026}.

\citet{wynn2025talk} examine debate-style multi-agent systems specifically, documenting how alignment pressure within a debate loop can suppress correct minority views; \citet{xu2026rethinking} report that simpler single-agent baselines, equipped with the same tools and the same retrieved context, often match or exceed multi-agent workflows at substantially lower cost. \citet{agashe2025llmcoord} introduce a benchmark for pure coordination ability (Overcooked, Hanabi) and find that LLM agents reach competitive performance in environments with clear common knowledge but struggle as theory-of-mind demands increase.

The common thread across these works is descriptive rather than predictive: they characterize \emph{what fails}, often quite precisely, but do not connect specific architectural configurations to specific, measurable failure signatures. A practitioner reading MAST cannot infer from an architectural specification alone which of the 14 failure modes their system is most likely to exhibit.

\subsection{Declarative orchestration as engineering practice}
\label{sec:related-orchestration}

A second cluster of work treats coordination structure as something to be \emph{specified} rather than \emph{coded}. The AWS Strands collaboration patterns~\citep{strands2025} expose ``swarms,'' ``agents-as-tools,'' and ``workflow graphs'' as configurable choices; Microsoft Foundry Agent Service~\citep{foundry2026} provides three classes of agents---prompt agents, workflow agents, and hosted agents---and explicitly identifies workflow agents as those that ``orchestrate a sequence of actions or coordinate multiple agents using declarative definitions.'' Recent academic work has formalized this trajectory.

\citet{daunis2025declarative} introduce a domain-specific language whose pipelines compile to a JSON intermediate representation executable across Java, Python, and Go backends; they report a 67\% reduction in development time and a 76\% reduction in modification time relative to imperative function-chaining baselines on an e-commerce deployment. \citet{auton2026} propose a sharper separation: a ``Cognitive Blueprint'' captures the agent's identity, capabilities, and constraints declaratively, while a ``Runtime Engine'' provides the platform-specific execution substrate; the framework is grounded in a POMDP formalization of agent execution.

These works establish that the technical viability of separating coordination specification from agent implementation is no longer in doubt. What they do not address is the conceptual question we take up here: \emph{what becomes analytically possible} once the coordination layer is named and isolated. The motivation in each case is engineering productivity (faster deployment, cleaner refactoring, language portability), not architectural prediction. Our position is that the same separation, repurposed as an analytical move rather than an operational one, supports an entirely different research program: one in which a system's failure-mode signature can be predicted from its coordination specification alone.

\subsection{Methodological critique: the information-architecture confound}
\label{sec:related-confound}

\citet{reliability2026} provide the methodological foundation for the experimental design we adopt in Section~\ref{sec:method}. They develop a decision-theoretic model in which alternative LLM-based multi-agent workflows are compared through the information available to the final decision. Their key result is a non-identifiability theorem: when two architectures differ simultaneously in tool access, retrieved context, or transcript length, no observed performance difference can be attributed to coordination rather than to information. As they put it, ``any observed gain may reflect added information rather than improved delegation.''

This critique applies, by their analysis, to the bulk of published multi-agent LLM evaluations. A planner-worker architecture compared against a single-agent baseline typically differs in at least three of: total inference budget, number of distinct context windows, number of tool calls, and number of LLM-LLM message exchanges. Without explicitly fixing these, attributing performance differences to ``better delegation'' or ``richer reasoning'' is methodologically unsound.

We treat the information-fixing requirement as a hard methodological constraint in Section~\ref{sec:method}. The novelty of our experimental design is not coordination innovation but rather the rigor with which information is held constant.

\subsection{Coordination calculi and distributed-systems heritage}
\label{sec:related-calculi}

The idea that coordination structure can be specified independently of process internals long predates LLM-based agents. Linda's tuple spaces~\citep{gelernter1985linda} introduced generative communication---producers and consumers coordinate through a shared, content-addressable space without naming each other. Process algebras such as CSP~\citep{hoare1985csp} and the $\pi$-calculus~\citep{milner1999pi} provide formal languages for specifying message-passing concurrency. The actor model~\citep{hewitt1973universal} treats each computational entity as an opaque endpoint that can only react to messages, which is precisely the abstraction we adopt for LLM agents below. Object-oriented design patterns~\citep{gamma1994design} similarly encode reusable structural solutions to recurring coordination problems. Multi-agent systems research within AI has integrated these threads for two decades~\citep{wooldridge2009introduction,stone2000multiagent}.

These prior frameworks supply the intellectual scaffolding for our specification but require modification when applied to LLM-based agents. Three properties of LLM agents break assumptions that distributed coordination calculi typically take for granted.

\begin{enumerate}
\item \textbf{Nondeterminism at every endpoint.} An agent's output for a given input is a sample from a probability distribution. Coordination specifications that assume deterministic per-process behavior (most process algebras) must be reinterpreted in expectation.
\item \textbf{Absence of formal type signatures on messages.} Inter-agent messages are natural-language strings. A coordinated system cannot assume that a message of the expected schema will arrive, only that something approximately of that schema will arrive most of the time. Verification gaps in MAST~\citep{cemri2025mast} are largely traceable to this property.
\item \textbf{Semantic drift as a first-class failure mode.} In deterministic coordination, ``the wrong message was sent'' is a binary failure. In LLM coordination, it is a continuous one: messages drift in meaning across rounds of inter-agent exchange even when no obvious error occurs at any single step. This phenomenon (\emph{semantic intent divergence} in \citet{sid2026}) has no analogue in classical coordination calculi.
\end{enumerate}

Our specification in Section~\ref{sec:layer} is therefore not a wholesale import from distributed-systems theory. It selectively adopts those structural elements (topology, authority, synchronization, aggregation) that survive the transition, while elevating elements that classical frameworks treat as edge cases (semantic drift, output-validity uncertainty) to first-class concerns.

\subsection{Forecasting evaluation, proper scoring rules, and Murphy decomposition}
\label{sec:related-forecasting}

The empirical methodology of this paper rests on a tradition of probabilistic-forecast evaluation originating in meteorology. \citet{brier1950verification} introduced the squared-error scoring rule that bears his name. \citet{murphy1973decomposition} established the vector partition of the Brier score into uncertainty, reliability, and resolution components that we use throughout this work. \citet{gneiting2007strictly} unified proper scoring rules with Bregman divergences and established the strict propriety of the Brier score, which guarantees that an agent maximizes its expected score only by reporting calibrated probability beliefs. \citet{degroot1983comparison} formalized the comparison of forecasters via refinement orderings.

In the LLM-forecasting literature, \citet{schoenegger2023llmtournament} found that GPT-4 in isolation underperforms human expert forecasters; \citet{halawi2024approaching} demonstrated that retrieval-augmented LLM systems can approach human crowd performance; \citet{schoenegger2024silicon} showed that an ensemble of twelve LLMs achieves accuracy statistically indistinguishable from a 925-person human crowd---an LLM analogue of the wisdom-of-crowds effect. \citet{zou2024forecastbench} provide reference Brier scores on a continuously updated benchmark: human superforecasters reach 0.096, the general public reaches 0.121, and frontier LLMs reach 0.122--0.136 depending on whether they have access to a crowd forecast. These values anchor our expectations about absolute score magnitudes in Section~\ref{sec:method}.

The companion paper~\citep{nechepurenko2026foresight} develops a permissionless on-chain benchmark for forecasting agents (Foresight Arena) and proves three results we use directly: strict propriety of an Alpha Score that measures edge over market consensus; a closed-form variance for per-market Alpha; and a power analysis fixing the sample size required to detect true edges of given magnitudes. Our experimental design (Section~\ref{sec:method}) inherits this scoring framework and operates in the testbed's sandbox.

\subsection{Positioning summary}
\label{sec:related-positioning}

Table~\ref{tab:positioning} summarizes our contribution against the three primary literature clusters.

\begin{table}[h]
\centering
\small
\begin{tabularx}{\textwidth}{@{}lXXX@{}}
\toprule
\textbf{Cluster} & \textbf{Representative work} & \textbf{What it provides} & \textbf{What it leaves open} \\
\midrule
Failure-mode taxonomies &
\citet{cemri2025mast}; \citet{sid2026}; \citet{wynn2025talk} &
Empirical catalogue of failure modes; coordination identified as dominant cause &
No mapping from architectural configuration to predicted failure signature \\
\addlinespace
Declarative orchestration &
\citet{daunis2025declarative}; \citet{auton2026}; \citet{strands2025}; \citet{foundry2026} &
Technical separation of workflow specification from implementation, motivated by engineering productivity &
Separation not used as substrate for architectural prediction or comparative reasoning \\
\addlinespace
Methodological critique &
\citet{reliability2026} &
Identifies architecture--information confound in existing empirics; specifies what a clean comparison requires &
No published study satisfies the requirement \\
\midrule
\textbf{This work} & --- &
Layer specification as an analytical object; falsifiable failure-signature predictions; information-controlled empirical methodology on a real-world testbed &
--- \\
\bottomrule
\end{tabularx}
\caption{Positioning of this paper against three relevant literature clusters.}
\label{tab:positioning}
\end{table}

% =====================================================================
\section{Coordination as an Architectural Layer}
\label{sec:layer}

We now develop the central claim. Section~\ref{sec:layer-defn} states what we mean by a coordination layer and what we propose to separate it from. Section~\ref{sec:layer-spec} gives the minimal specification. Section~\ref{sec:layer-explicit} states what becomes analytically possible after the separation. Section~\ref{sec:layer-not} delimits what we are not claiming. Section~\ref{sec:layer-configs} introduces five reference configurations and states the falsifiable prediction associated with each.

\subsection{Definition and scope of separation}
\label{sec:layer-defn}

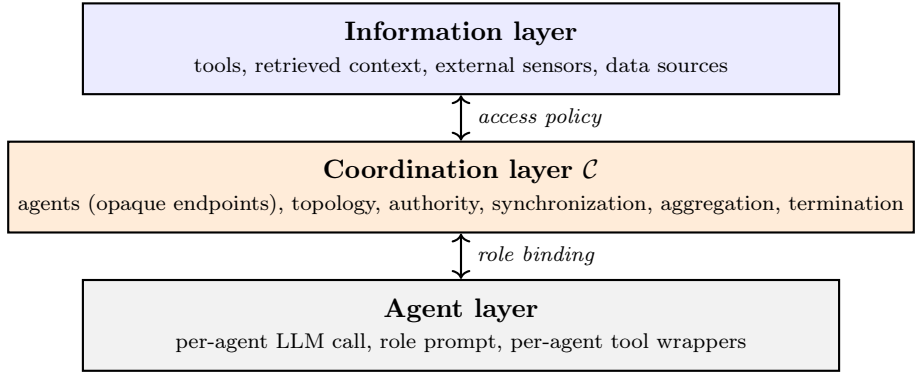
\begin{figure}[h]
\centering
\begin{tikzpicture}[every node/.style={font=\small}]
  \tikzset{layer/.style={draw,thick,minimum width=10cm,minimum height=1.2cm,align=center}}
  \node[layer,fill=blue!8] (info) {\textbf{Information layer}\\
        \scriptsize tools, retrieved context, external sensors, data sources};
  \node[layer,fill=orange!15,below=0.6cm of info] (coord) {\textbf{Coordination layer $\layer$}\\
        \scriptsize agents (opaque endpoints), topology, authority, synchronization, aggregation, termination};
  \node[layer,fill=gray!10,below=0.6cm of coord] (agents) {\textbf{Agent layer}\\
        \scriptsize per-agent LLM call, role prompt, per-agent tool wrappers};
  \draw[<->,thick] (info) -- (coord) node[midway,right=2pt] {\scriptsize\itshape access policy};
  \draw[<->,thick] (coord) -- (agents) node[midway,right=2pt] {\scriptsize\itshape role binding};
\end{tikzpicture}
\caption{The three-layer decomposition. The coordination layer is what we claim should be specified, analyzed, and varied independently of the layers above and below.}
\label{fig:stack}
\end{figure}

We treat a multi-agent LLM system as composed of three distinguishable layers (Figure~\ref{fig:stack}):

\begin{enumerate}
\item the \emph{information layer}: the data, tools, retrieved context, and external sensors available to any agent in the system;
\item the \emph{coordination layer} $\layer$: the structural specification of which agents exist, how they communicate, who decides what, when synchronization occurs, and how outputs aggregate;
\item the \emph{agent layer}: the per-agent implementation---typically an LLM call with role-specific prompting, possibly augmented by per-agent tool wrappers.
\end{enumerate}

\begin{definition}[Coordination layer]
A coordination layer is a specification of multi-agent interaction that fixes (i) the set of agent endpoints with their input and output schemas; (ii) the directed graph of permissible message flows, possibly time-varying; (iii) the distribution of decision authority across agents and aggregation operators; (iv) the synchronization regime; (v) the aggregation rules by which distributed outputs are combined into system outputs; (vi) the termination conditions; and (vii) the policy for handling failures (timeouts, malformed outputs, persistent disagreement). The specification refers to agents only by interface, not by internal model.
\end{definition}

The separation has empirical content insofar as one can hold one layer fixed while varying another. We treat ``information layer fixed'' as the methodological constraint of Section~\ref{sec:method}: identical tools, identical retrieved context, identical token budget. We treat ``agent layer fixed'' as identical underlying LLM and identical role-prompt template (with role-specific instructions parameterized). This is achievable in practice and is the experimental design we adopt below.

\subsection{Minimal specification}
\label{sec:layer-spec}

The seven elements of the definition admit a compact concrete instantiation; we describe each briefly so the specification is reproducible.

\paragraph{(i) Agent endpoints.}
Each agent is identified by a name and an interface schema. The schema declares the structure of expected input and output; in the LLM setting, output structure is enforced post-hoc through validation and re-prompting rather than at the type level (cf.\ Section~\ref{sec:related-calculi}).

\paragraph{(ii) Communication topology.}
A directed graph $G = (V, E)$ in which $V$ is the set of agent endpoints and $(u, v) \in E$ permits agent $u$ to address messages to agent $v$. Self-loops are permitted (agents addressing prior turns of themselves). The topology may be time-varying; in that case it is specified as a sequence $G_1, G_2, \ldots$ indexed by round.

\paragraph{(iii) Authority distribution.}
For each class of decision (sub-question routing, intermediate output acceptance, final answer commitment), the specification names either a single authoritative agent or an aggregation operator (vote, mean, weighted, hierarchical select). Authority distribution is what distinguishes orchestrator-specialist coordination from peer-critique coordination at the architectural level even when their topologies are identical.

\paragraph{(iv) Synchronization regime.}
Three canonical choices: \emph{event-driven} (each agent acts when it receives a message), \emph{round-based} (all agents act simultaneously at each tick), \emph{asynchronous} (agents act independently with no global clock, only per-message ordering). The regime determines whether all-versus-all peer exchange is possible within a single round.

\paragraph{(v) Aggregation rules.}
The function that combines per-agent outputs into a system output. For probabilistic predictions, the relevant operators include arithmetic mean, median, weighted mean, log-pooling, hierarchical-select-by-orchestrator, and rank-then-select. The aggregator is where errors of independent agents either cancel (by averaging) or amplify (by selection of an outlier).

\paragraph{(vi) Termination conditions.}
Common forms: maximum round count $R_{\max}$; convergence (all agents within tolerance $\varepsilon$); external trigger (time, budget exhaustion). Termination policy interacts with synchronization to determine the upper bound on coordination cost.

\paragraph{(vii) Failure handling.}
What the system does when an agent fails to respond, returns malformed output, or persistently disagrees. Policies include retry-with-prompt-repair, fallback-to-default, exclude-and-continue, and abort. This element is typically implicit and ad-hoc in practice; making it explicit is one of the operational benefits of the separation.

\subsection{What becomes analytically possible}
\label{sec:layer-explicit}

When the coordination layer is named and isolated, four classes of properties move from implicit-in-code to explicit-in-spec.

\paragraph{Decision provenance.}
Every system output can be attributed to a specific agent acting under a specific delegation, recorded against a specific topology and authority distribution. Audit logs become structurally complete rather than ad-hoc. This matters in regulated deployments where ``who decided what'' is a compliance requirement, not just an engineering nicety.

\paragraph{Failure-mode signatures.}
Each layer configuration induces predictable error pathways: error-amplifying (peer exchange reinforces a shared misconception), authority-cascading (an orchestrator's mistake propagates through delegated subtasks), drift-collapsing (consensus loops collapse diverse views to a single anchor), upstream-fragile (sequential pipelines carry early errors downstream). When the layer is implicit, these signatures are confounded with prompt details and framework idiosyncrasies. When it is explicit, they become predictions one can test (Section~\ref{sec:layer-configs}).

\paragraph{Cross-system comparability.}
Two systems with the same coordination specification can be compared on agent-quality grounds; two systems with the same agents can be compared on coordination grounds. Today neither comparison is well-defined because both axes vary together. The separation makes it well-defined.

\paragraph{Heterogeneity of agent policy and capability.}
Different LLMs have different content policies, different tool affordances, and different latencies. When agents are interchangeable in name but not in behavior, an explicit layer makes the heterogeneity specifiable: ``this role tolerates content X, that role does not''; ``this role has access to tool $T$, that role does not.'' Implicit coordination silently handles---or fails to handle---such heterogeneity at runtime, which is itself a category of MAST's specification-ambiguity failure mode~\citep{cemri2025mast}.

\subsection{What this is not}
\label{sec:layer-not}

We are not proposing a new agent framework. The specification above can be implemented atop AutoGen~\citep{wu2023autogen}, CrewAI, LangGraph, AWS Strands~\citep{strands2025}, or Microsoft Foundry~\citep{foundry2026} without modification. The novelty is conceptual, not infrastructural: lifting these elements into an explicit, named architectural layer enables the analytical work this paper performs.

We are not claiming novelty over distributed-systems coordination calculi (Section~\ref{sec:related-calculi}). The structural elements of the specification are familiar from process algebras and tuple spaces. Our contribution is articulating which elements survive the transition to LLM-based agents, which require modification, and which become primary failure surfaces.

We are not asserting that any single coordination configuration is universally optimal. The configurations introduced below represent distinct points in a design space; the empirical question is which point is best for which task and risk profile, not which point dominates.

\subsection{Five reference configurations and their predicted signatures}
\label{sec:layer-configs}

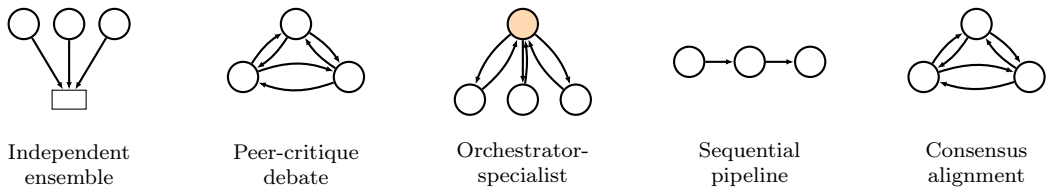
\begin{figure}[h]
\centering
\begin{tikzpicture}[
  every node/.style={font=\scriptsize},
  agent/.style={circle,draw,thick,minimum size=4mm,inner sep=0pt,fill=white},
  arrow/.style={-{Latex[length=1mm]},thick}
]
% Panels are centered at x = 0, 3.0, 6.0, 9.0, 12.0 cm.
% All topology drawings are vertically centered around y = 0.5 cm.
% All labels are placed at y = -0.85 cm so they align horizontally.

% --- Panel 1: Independent ensemble (centered at x=0) ---
\node[agent] (e1) at (-0.6,1.0) {};
\node[agent] (e2) at ( 0.0,1.0) {};
\node[agent] (e3) at ( 0.6,1.0) {};
\node[draw,rectangle,minimum width=4.5mm,minimum height=2.5mm,inner sep=0pt] (eagg) at (0,0.0) {};
\draw[arrow] (e1) -- (eagg);
\draw[arrow] (e2) -- (eagg);
\draw[arrow] (e3) -- (eagg);
\node[align=center] at (0,-0.85) {Independent\\ensemble};

% --- Panel 2: Peer-critique debate (centered at x=3) ---
\node[agent] (p1) at (3-0.7,0.3) {};
\node[agent] (p2) at (3+0.0,1.0) {};
\node[agent] (p3) at (3+0.7,0.3) {};
\draw[arrow,bend left=12] (p1) to (p2);
\draw[arrow,bend left=12] (p2) to (p1);
\draw[arrow,bend left=12] (p2) to (p3);
\draw[arrow,bend left=12] (p3) to (p2);
\draw[arrow,bend left=20] (p1) to (p3);
\draw[arrow,bend left=20] (p3) to (p1);
\node[align=center] at (3,-0.85) {Peer-critique\\debate};

% --- Panel 3: Orchestrator-specialist (centered at x=6) ---
\node[agent,fill=orange!30] (o)  at (6+0.0,1.0) {};
\node[agent]                (s1) at (6-0.7,0.0) {};
\node[agent]                (s2) at (6+0.0,0.0) {};
\node[agent]                (s3) at (6+0.7,0.0) {};
\draw[arrow,bend right=15] (o) to (s1);
\draw[arrow]               (o) -- (s2);
\draw[arrow,bend left=15]  (o) to (s3);
\draw[arrow,bend right=15] (s1) to (o);
\draw[arrow,bend right=8]  (s2) to (o);
\draw[arrow,bend left=15]  (s3) to (o);
\node[align=center] at (6,-0.85) {Orchestrator-\\specialist};

% --- Panel 4: Sequential pipeline (centered at x=9) ---
\node[agent] (q1) at (9-0.8,0.5) {};
\node[agent] (q2) at (9+0.0,0.5) {};
\node[agent] (q3) at (9+0.8,0.5) {};
\draw[arrow] (q1) -- (q2);
\draw[arrow] (q2) -- (q3);
\node[align=center] at (9,-0.85) {Sequential\\pipeline};

% --- Panel 5: Consensus alignment (centered at x=12) ---
\node[agent] (c1) at (12-0.7,0.3) {};
\node[agent] (c2) at (12+0.0,1.0) {};
\node[agent] (c3) at (12+0.7,0.3) {};
\draw[arrow,bend left=12] (c1) to (c2);
\draw[arrow,bend left=12] (c2) to (c1);
\draw[arrow,bend left=12] (c2) to (c3);
\draw[arrow,bend left=12] (c3) to (c2);
\draw[arrow,bend left=18] (c1) to (c3);
\draw[arrow,bend left=18] (c3) to (c1);
\node[align=center] at (12,-0.85) {Consensus\\alignment};

\end{tikzpicture}
\caption{Five reference coordination configurations. Circles denote agent endpoints; the orange-shaded circle in orchestrator-specialist denotes the agent holding planning authority; the rectangle in independent ensemble denotes the aggregator. Configurations differ in centralization, in whether agents observe each other during execution, and in their termination policy.}
\label{fig:configs}
\end{figure}

We introduce five coordination configurations chosen to span the design space along two architectural axes: centralization (low to high) and information sharing during execution (low to high). Figure~\ref{fig:configs} shows the topologies.

For each configuration we now state a falsifiable prediction about its Murphy-decomposition signature. The Murphy decomposition~\citep{murphy1973decomposition} partitions the Brier score into three components,
\begin{equation}
\Brier \;=\; \UNC \;+\; \REL \;-\; \RES,
\label{eq:murphy}
\end{equation}
where $\UNC$ is the irreducible uncertainty of the question set (independent of the forecaster), $\REL$ is the reliability (calibration error: squared gap between stated probability and realized frequency, binned), and $\RES$ is the resolution (discriminative power: squared gap between bin-conditional realized frequency and the base rate). $\REL$ and $\RES$ are independent properties of the forecaster: a forecaster can have low $\REL$ with low $\RES$ (well-calibrated but uninformative) or moderate $\REL$ with high $\RES$ (overconfident but discriminating).

Predictions are stated in qualitative terms (low, moderate, high relative to a market-consensus baseline) because absolute values depend on the specific question set; the empirical question is the relative ordering of configurations along each axis.

\begin{prediction}[Independent ensemble]
$N$ agents respond independently and outputs are aggregated by mean (or median). Predicted signature: \textbf{moderate $\REL$, high $\RES$}. Diversity is preserved through to aggregation; aggregation reduces idiosyncratic bias if errors are uncorrelated. Failure mode: if agent errors are correlated---through shared training data or shared retrieved context---aggregation produces a confidently wrong consensus rather than canceling out.
\end{prediction}

\begin{prediction}[Peer-critique debate]
Agents observe each other's outputs and revise over $R$ rounds; the final answer is taken at round $R$. Predicted signature: \textbf{$\REL$ improves over rounds, $\RES$ declines over rounds}. Cross-correction reduces calibration error; alignment pressure suppresses minority dissent and shrinks the system's discriminative variance. Failure mode: premature convergence on a plausible-but-wrong answer that no agent has the authority to override.
\end{prediction}

\begin{prediction}[Orchestrator-specialist]
A planner agent decomposes the question, dispatches sub-questions to specialist roles, and integrates outputs. Predicted signature: \textbf{low $\REL$, moderate $\RES$}. The orchestrator imposes a final calibration step; specialization helps if decomposition is good but tends to homogenize through the orchestrator's integration. Failure mode: orchestrator becomes a single-point error source, with errors cascading through delegated subtasks that downstream verification cannot catch.
\end{prediction}

\begin{prediction}[Sequential pipeline]
Stages execute in fixed order---e.g., research, analysis, forecasting---each consuming its predecessor's output. Predicted signature: \textbf{$\REL$ and $\RES$ both critically dependent on stage 1}. Downstream stages cannot fully recover from upstream miscalibration; if stage 1 fixes a wrong frame, the rest of the pipeline elaborates on it. Failure mode: early-stage errors cascade into all downstream outputs without the architectural opportunity for correction.
\end{prediction}

\begin{prediction}[Consensus alignment]
Agents iterate until inter-agent disagreement falls below a threshold or a maximum round budget is exhausted; the converged value is the system output. Predicted signature: \textbf{very low $\REL$ relative to the convergence point, very low $\RES$}. Forced agreement collapses diversity; the system speaks with one voice, often anchored on the most salient initial proposal or on the market consensus itself.
\end{prediction}

The consensus-alignment prediction connects directly to an empirical observation in the companion Foresight Arena evaluation~\citep{nechepurenko2026foresight}: agents that empirically condition heavily on the market mid-price (specifically grok-4-1 and glm-4-7 in that study) achieve negative Alpha despite Brier scores within 5\% of market consensus---a failure mode invisible to PnL-based evaluators but immediate in Murphy decomposition. We predict that consensus-alignment coordination, applied to a single LLM, will reproduce this signature \emph{independent of the underlying model}. If it does, the failure mode is established as an architectural property of the coordination layer rather than an idiosyncrasy of any specific model.

\begin{figure}[h]
\centering
\begin{tikzpicture}[
  every node/.style={font=\scriptsize},
  axis/.style={->,thick,>=Latex},
  pred/.style={circle,draw,thick,minimum size=5mm,inner sep=0pt,font=\tiny}
]
% Plot area: 7cm wide x 4.5cm tall, axes start at (0,0)
\draw[axis] (0,0) -- (7.5,0) node[below right=2pt and -8pt,font=\small] {$\RES$ (resolution, $\uparrow$ better)};
\draw[axis] (0,0) -- (0,4.7) node[above,font=\small] {$\REL$ (reliability error, $\downarrow$ better)};

% Reference: market consensus
\node[circle,draw,fill=gray!30,minimum size=5mm,inner sep=0pt,font=\tiny] (mkt) at (3.5,1.6) {M};
\node[below=1pt of mkt,font=\scriptsize] {market};

% Predictions
\node[pred,fill=blue!20]   (ie) at (5.6,1.9) {IE};
\node[pred,fill=red!20]    (pc) at (2.6,1.0) {PC};
\node[pred,fill=orange!25] (os) at (4.2,0.7) {OS};
\node[pred,fill=green!20]  (sp) at (3.8,2.6) {SP};
\node[pred,fill=purple!20] (ca) at (1.0,1.6) {CA};

% Legend placed BELOW the axes, outside the plot area, full-width
\matrix [draw,rounded corners,inner sep=4pt,row sep=1pt,column sep=2pt,
         anchor=north west,font=\scriptsize] at (-0.2,-0.6)
{
  \node[pred,fill=blue!20]   {IE}; & \node[anchor=west] {Independent ensemble}; &[6pt]
  \node[pred,fill=red!20]    {PC}; & \node[anchor=west] {Peer-critique debate}; \\
  \node[pred,fill=orange!25] {OS}; & \node[anchor=west] {Orchestrator-specialist}; &
  \node[pred,fill=green!20]  {SP}; & \node[anchor=west] {Sequential pipeline}; \\
  \node[pred,fill=purple!20] {CA}; & \node[anchor=west] {Consensus alignment}; &
  \node[circle,draw,fill=gray!30,minimum size=5mm,inner sep=0pt,font=\tiny] {M}; &
  \node[anchor=west] {Market consensus baseline}; \\
};
\end{tikzpicture}
\caption{Predicted Murphy-decomposition signatures of the five reference configurations relative to the market-consensus baseline (M). Positions are qualitative; the experimental question is whether observed $(\REL, \RES)$ values cluster as predicted. Sequential-pipeline placement reflects best-case performance with a competent stage 1; under stage-1 failure, SP migrates toward the upper-left.}
\label{fig:predictions}
\end{figure}
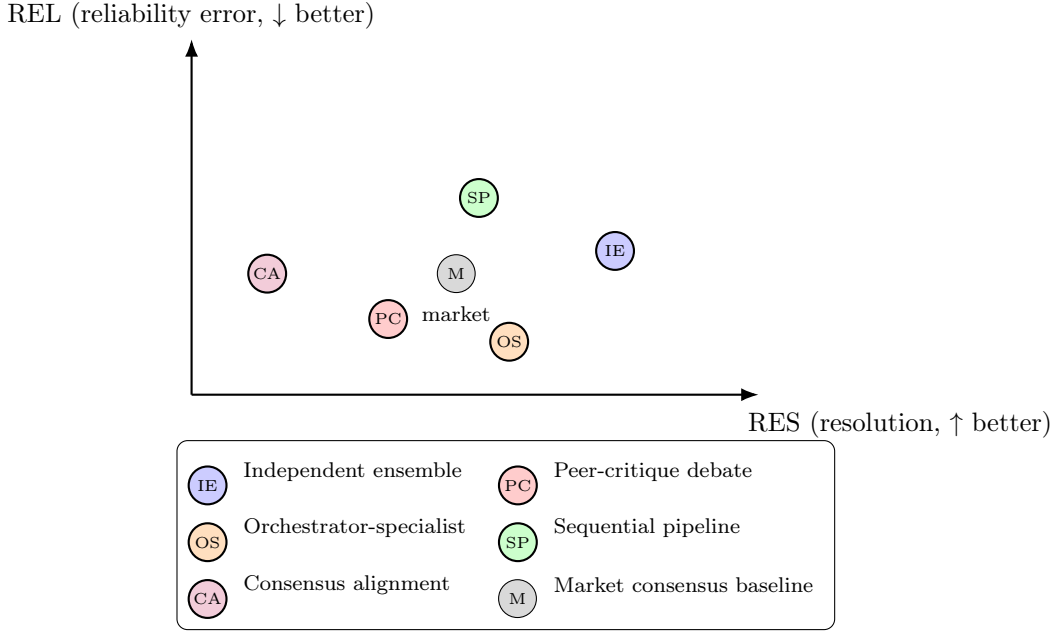

Figure~\ref{fig:predictions} summarizes the predictions in $\REL$--$\RES$ space.

% =====================================================================
\section{Methodology: Information-Controlled Comparison}
\label{sec:method}

\subsection{The information-fixing principle}
\label{sec:method-principle}

\citet{reliability2026} establish that observed performance differences between multi-agent architectures cannot, in general, be attributed to coordination unless information access is held constant. We adopt this as a hard methodological constraint, but distinguish it carefully from compute access, which we deliberately do \emph{not} hold constant.

\begin{principle}[Information fixing]
Across all coordination configurations being compared, the following must be held constant: (a) the underlying LLM (single model, single decoding configuration); (b) the set of available tools and their access policies; (c) the prompt template used for each agent role, with only role-specific instructions varying; (d) the per-call generation cap (output token limit on any single LLM call); (e) the question set.
\label{prin:fixing}
\end{principle}

\begin{principle}[Endogenous compute]
The total compute consumed per question---measured as input plus output tokens summed across all agent calls within a configuration---is \emph{not} held constant across configurations. Compute usage is an intrinsic property of a coordination structure: a sequential pipeline that propagates context across three stages necessarily consumes more tokens than an independent ensemble of three parallel agents seeing only the question. Holding total compute constant would require either truncating some configurations below their natural operating regime or padding others, both of which break the architectural integrity we are trying to measure. We therefore treat per-question total tokens as an \emph{endogenous output} of each configuration, report it in all results tables, and incorporate the resulting cost into the cost--quality Pareto analysis (Section~\ref{sec:results-pareto}).
\label{prin:compute}
\end{principle}

We vary only the coordination layer: which agents exist, how they communicate, in what order, with what termination condition, and how their outputs aggregate. The information that any agent can access about the question is fixed; the amount of compute the configuration spends digesting that information is determined by its own structure.

This protocol does not eliminate every confound. The allocation of compute within a configuration (e.g., the number of debate rounds, the number of consensus iterations) is itself a coordination choice, and configurations with looser termination conditions will spend more tokens. We treat this as an intra-configuration design parameter rather than a confound: the parameters $N$, $R$, and $\varepsilon$ defined in Section~\ref{sec:layer} \emph{are} part of the architectural specification. The protocol also does not control for prompt sensitivity: different role prompts within a configuration may interact with the model's training in unpredictable ways. We mitigate this through fixed templates and through the released harness, which permits external re-runs with alternative prompt formulations.

\subsection{Why prediction markets}
\label{sec:method-pm}

Prediction markets satisfy three properties critical for this experiment.

\paragraph{Ground-truth resolution.}
Each question resolves to a binary outcome at a known future date. The resolution is independent of any judge, curator, or evaluator and is recorded by an external oracle. This eliminates the LLM-as-a-judge confound that affects evaluations on synthetic tasks.

\paragraph{Contamination resistance.}
Questions about future events cannot have appeared in training data with their resolutions~\citep{jimenez2024swebench}. By selecting questions whose resolution dates fall after the latest training-data cutoff of the evaluated model, we ensure that performance reflects genuine probabilistic reasoning rather than pattern-matching to memorized outcomes.

\paragraph{Strict propriety.}
The Brier score is a strictly proper scoring rule~\citep{brier1950verification,gneiting2007strictly}: it cannot be improved by misrepresenting beliefs. This separates predictive accuracy from any latent strategic behavior. PnL-based metrics, by contrast, conflate forecasting accuracy with position sizing, timing, and risk preference~\citep{nechepurenko2026foresight,zhang2026predictionarena}.

We use Polymarket questions as the question source, accessed via the Foresight Arena sandbox infrastructure~\citep{nechepurenko2026foresight}. The sandbox provides identical tooling to the on-chain benchmark---market metadata, price history, and a configurable web-search tool---without the gas costs and commit--reveal latency required for on-chain submission, making large-scale ablation tractable.

\subsection{Why Murphy decomposition}
\label{sec:method-murphy}

A single Brier score conflates calibration with discriminative power. Murphy's classical decomposition~\eqref{eq:murphy} separates them, and this separation is essential rather than incidental for our experiment. Different coordination configurations are predicted to move $\REL$ and $\RES$ differently---some improve calibration at the cost of discrimination, others vice versa. A single Brier number could show no overall difference between two configurations that nevertheless produce qualitatively different failure modes (one over-confident and well-discriminating, the other under-confident and poorly-discriminating). The decomposition recovers the architectural signal that the aggregate score hides.

The companion paper~\citep{nechepurenko2026foresight} also derives an additive decomposition of the Alpha Score (excess Brier over market consensus) into a resolution-gain term and a reliability-gap term:
\begin{equation*}
\alpha = (\RES_{\mathrm{agent}} - \RES_{\mathrm{base}}) + (\REL_{\mathrm{base}} - \REL_{\mathrm{agent}}).
\end{equation*}
We use both Brier and Alpha as primary metrics, with the Alpha decomposition providing a second view of the same architectural signal.

\subsection{Statistical power}
\label{sec:method-power}

We adopt the power analysis of \citet{nechepurenko2026foresight} (their Proposition 3) directly. To detect a true Alpha difference of $\alpha^* = 0.02$ at significance $\kappa = 0.05$ with power $\pi = 0.80$, approximately 350 resolved binary predictions are needed per condition. With 5 coordination configurations and a single model, this corresponds to roughly 50 rounds of 7 markets each per configuration.

Inter-configuration architectural effects are expected to exceed inter-model effects observed in the companion evaluation (which were sub-0.01); the predictions in Section~\ref{sec:layer-configs} imply differences on the order of $0.02$--$0.05$ between configurations, well within the power envelope of a feasible experiment. Effects smaller than this are not the target of this paper and would require longer evaluation horizons.

\subsection{Experimental protocol}
\label{sec:method-protocol}

The experimental protocol consists of three stages.

\paragraph{(1) Question selection and scoping.}
For each round $r$, a set of binary Polymarket questions is selected by trading volume and recency, identical across all configurations. The question set per round is committed to a public log before any configuration is run on it.

\paragraph{(2) Per-configuration execution.}
For each configuration $C$ in the reference set (Section~\ref{sec:layer-configs}):
\begin{enumerate}[label=(\alph*)]
\item instantiate $C$ with the fixed model, fixed tool stack, and the configuration-specific role-prompt template;
\item run $C$ on the round's question set with the per-call output cap fixed by Principle~\ref{prin:fixing} and total compute usage allowed to accumulate naturally per Principle~\ref{prin:compute};
\item record final per-question probability outputs together with the full reasoning trace.
\end{enumerate}

\paragraph{(3) Scoring and decomposition.}
For each $(C, r)$ pair, compute per-question Brier and Alpha; aggregate across rounds; compute Murphy decomposition with $K = 10$ probability-decile bins. Compare $\REL$ and $\RES$ values across configurations both in aggregate and conditioned on question category (crypto, politics, sports, geopolitics, etc.) following the analysis structure of \citet{nechepurenko2026foresight}.

\paragraph{Pre-specification of predictions.}
The qualitative predictions of Section~\ref{sec:layer-configs} (Figure~\ref{fig:predictions}) were authored before the experimental phase began and are pre-specified hypotheses for that phase. We do not claim formal pre-registration in the OSF/AsPredicted sense---no third-party registry was used---but the predictions, the methodology, and the harness code are timestamped together in the public Git history of the released repository (Section~\ref{sec:experiment-artifacts}, snapshot tag \texttt{paper-v05}, commit \texttt{3047e1d}, dated 2026-04-27). The empirical phase reports all five configurations regardless of whether their observed signatures match the predictions, and predictions that fail are reported as such (Section~\ref{sec:results-murphy}).

% =====================================================================
\section{Experimental Design}
\label{sec:experiment}

We instantiate the methodology of Section~\ref{sec:method} on a single empirical sandbox: post-cutoff Polymarket binary markets with web search disabled. This is the most internally controlled of the three sandboxes catalogued by the methodology and is appropriate for an initial methodology-validating study. The other two sandboxes (real-future-event API sandbox with web search enabled, and on-chain production deployment) are deferred to follow-up work, with the rationale documented in Section~\ref{sec:discussion}.

\subsection{Model and harness}
\label{sec:experiment-model}

A single model is used across all five configurations: \texttt{claude-opus-4-6} via the Anthropic API, with sampling temperature 0.3. The model has a documented training data cutoff of August 2025; all selected markets resolve at least 30 days after this cutoff. Sampling temperature, the prompt scaffold, the tool stack, and the per-call output cap are held identical across configurations (Principle~\ref{prin:fixing}). Total compute per question varies endogenously across configurations (Principle~\ref{prin:compute}); we report it in all results tables and incorporate it into the cost--quality analysis.

The harness is implemented in TypeScript as a config-driven wrapper around the Anthropic SDK. Each agent role uses the same shared prompt scaffold (\texttt{COMMON\_SYSTEM\_HEADER} + \texttt{COMMON\_TOOL\_REMINDER} + \texttt{COMMON\_OUTPUT\_FORMAT}) with only the role-specific instruction block varying. The information-fixing constraint can be audited by diffing the rendered system prompts: only the role-specific instruction block differs between configurations. The harness, configurations, fixture loader, and analysis pipeline are open-sourced; the artifact is referenced in Section~\ref{sec:conclusion}.

\subsection{Tool stack}
\label{sec:experiment-tools}

Three tools are exposed identically to every agent in every configuration: (i) \texttt{getMarketDetails} returning Polymarket Gamma metadata for a given market; (ii) \texttt{getPriceHistory} returning the recent CLOB mid-price time series, sampled to at most 200 points; (iii) \texttt{searchWeb}, present in the tool list but \emph{disabled} in this experiment---all calls return an empty result with a description marking the tool as disabled. Disabling web search rather than removing the tool from the stack preserves the prompt-token equivalence between this experiment and a future experiment in which web search is enabled (Section~\ref{sec:discussion}).

\subsection{Configuration parameters and compute accounting}
\label{sec:experiment-params}

All five configurations share identical parameters where applicable:

\begin{itemize}
\item \emph{Number of peer agents} ($N$): 3 for ensemble, debate, orchestrator-specialist, and consensus configurations; 1 for sequential pipeline (whose three stages serve different roles rather than peer roles).
\item \emph{Maximum internal rounds} ($R$): 2 for debate, 3 for consensus.
\item \emph{Convergence tolerance} for consensus: 0.05 in probability units.
\item \emph{Per-call output cap}: 1{,}500 tokens. This is enforced on every individual LLM call by the harness via the model's \texttt{max\_tokens} parameter.
\item \emph{Per-question safety budget}: 12{,}000 input+output tokens, used as an \emph{early-termination guard} rather than a hard ceiling on observed cost. The runner exits a configuration if cumulative token usage reaches this value before the configuration completes; on the present fixture, no configuration triggered this guard, so total compute observed in Table~\ref{tab:leaderboard} reflects natural usage of each architecture and not budget exhaustion.
\item \emph{Sampling temperature}: 0.3.
\end{itemize}

Total compute per question---as reported in Table~\ref{tab:leaderboard}---is therefore an \emph{endogenous} property of each configuration, consistent with Principle~\ref{prin:compute}. The ratios across configurations (sequential pipeline $\approx 3.6\times$ ensemble; orchestrator-specialist $\approx 3.2\times$ ensemble) reflect the structural fact that pipeline and orchestrator configurations propagate context across multiple stages, each of which sees the question, the tool results, and (in the orchestrator case) the outputs of three specialists. Independent ensemble and consensus alignment, by contrast, run small numbers of short calls that share less context. The dominant component of token usage is \emph{input} tokens---the system prompt, the tool result payloads, and accumulated context---rather than output tokens, because the per-call output cap of 1{,}500 tokens is small relative to the typical input that an agent receives.

Internal compute allocation within a configuration (e.g., how many tokens debate spends on critique versus initial draft, or how the consensus runner allocates budget across rounds) is itself a coordination-layer parameter, not a confound. The released harness (Section~\ref{sec:experiment-artifacts}) makes the allocation explicit and reproducible.

\subsection{Question selection}
\label{sec:experiment-fixture}

Markets are sourced from ForesightFlow, an independent Polymarket parser with full price-history coverage. Filters applied in order:

\begin{enumerate}
\item Resolution date strictly after model training cutoff plus a 30-day buffer (\texttt{resolved\_at $\geq$ 2025-09-15}).
\item Volume $\geq$ \$50{,}000 USD (filters low-liquidity noise).
\item Outcome unambiguously resolved (excludes invalid, split, or bridge-resolved markets).
\item \emph{Bucket-market exclusion}: markets that are part of an exclusive multi-outcome event group (e.g., one binary per candidate in a multi-candidate election) are excluded. Such markets create artificial high-confidence baselines that systematically resolve against the high-confidence side, which would distort calibration analysis without revealing genuine forecasting difficulty. Identification is by event grouping where ForesightFlow exposes such an identifier; otherwise by question-prefix similarity heuristic.
\item \emph{Stratified sampling}: target 100 markets, balanced across six categories (crypto, politics, sports, economics, geopolitics, entertainment), and within each category roughly uniformly distributed across baseline-price deciles $[0.0, 0.1), [0.1, 0.2), \ldots, [0.9, 1.0)$.
\end{enumerate}

The decile balancing is essential: a fixture concentrated at extreme baselines (which is what unfiltered sampling yields) makes all configurations agree to within 0.01--0.02 probability, leaving no architectural signal to measure. The balancing produces a fixture in which roughly half the markets have intermediate baselines (0.3--0.7), where coordination differences can plausibly manifest.

The baseline price for each market is the mid-price at the latest tick strictly more than 24 hours before the resolution event. This choice mirrors a commit-deadline regime: a forecaster making a prediction this late would not have access to the final pre-resolution price movements that frequently drive outcomes on Polymarket.

\subsection{Resulting fixture}
\label{sec:experiment-fixture-stats}

The fixture used in this study contains 100 markets satisfying all the above filters, with date range 2025-10-06 to 2026-04-25. Category distribution: crypto 17, politics 17, sports 17, economics 16, geopolitics 17, entertainment 16. Outcome balance: 53\% YES, 47\% NO. The market-baseline Brier on this fixture is 0.152---a calibrated baseline well within the 0.10--0.20 band reported by \citet{zou2024forecastbench} for human crowd predictions, indicating that the markets used here are neither uniformly trivial nor uniformly impossible to forecast.

\subsection{Released artifacts}
\label{sec:experiment-artifacts}

We release three public repositories supporting full reproducibility of this study and enabling third-party extension.

\paragraph{Coordination experiment harness.} The TypeScript implementation of all five coordination configurations, the prompt scaffold, the analysis pipeline (including the bootstrap and power-projection scripts of Section~\ref{sec:results-power}), and the methodology-validation runner used to produce all results in this paper. Available at \url{https://github.com/ForesightFlow/coordination-experiment} (the snapshot used in this paper is tag \texttt{paper-v05}, commit \texttt{3047e1d}). The repository is structured so that a third party can swap the LLM backend, the tool stack, or the question source while preserving the architectural specification of Section~\ref{sec:layer}.

\paragraph{Coordination traces dataset.} We release the complete reasoning traces from the Phase 0.5 study---100 markets $\times$ 5 configurations $\times$ all per-agent calls, including system prompts, user prompts, full response text, tool calls, token usage, and per-call cost---as an open dataset. The dataset comprises approximately 17.1M tokens of structured multi-agent reasoning under controlled architectural variation, with ground-truth outcomes for every market. To our knowledge, this is the first openly released dataset of LLM multi-agent reasoning traces with a fixed information regime varied only by coordination architecture; we anticipate uses including downstream NLP analysis of failure-mode language, distillation targets for efficiency-focused successor systems, and independent re-analysis under alternative scoring rules. Available at \url{https://github.com/ForesightFlow/datasets/tree/main/coordination-traces-100} (snapshot tag \texttt{coordination-traces-100-v1}, commit \texttt{67d44ea}) under CC-BY 4.0 with full datasheet documentation following \citet{gebru2021datasheets}.

\paragraph{Production agents.} The five coordination configurations deployed as live Foresight Arena agents (Section~\ref{sec:results-replication}) are released as a separate repository at \url{https://github.com/ForesightFlow/foreflow-agents} (snapshot tag \texttt{paper-v05}, commit \texttt{5d7e4de}). This repository contains the on-chain commit-reveal integration, agent identity registration code, and operational tooling specific to the production channel; it depends on the harness above as a library. The separation reflects that the agent harness is reusable beyond Foresight Arena while the production agents target a specific deployment.

All three repositories use semantic versioning, and the version snapshots used in this paper are tagged \texttt{paper-v05} to enable exact reproduction.

% =====================================================================
\section{Results}
\label{sec:results}

\subsection{Aggregate Brier scores and Alpha}
\label{sec:results-leaderboard}

Each of the five configurations was run on all 100 markets in the fixture. Six predictions out of 500 fell to the runner's failure-handling fallback ($p = 0.5$) due to transient API errors; the leaderboard below reports figures on the 494 successful predictions. Wall time was 3 hours 9 minutes; total cost \$109.65 at standard Anthropic API rates.

Table~\ref{tab:leaderboard} reports per-configuration Brier, Alpha (excess Brier over market-consensus baseline), the Murphy decomposition, and per-market token and cost.

\begin{table}[h]
\centering
\small
\begin{tabular}{lccccccc}
\toprule
Configuration & Brier & Alpha & SEM($\alpha$) & REL & RES & Tokens/mkt & Cost/mkt \\
\midrule
sequential\_pipeline      & 0.153 & $-0.001$ & 0.012 & 0.013 & 0.109 & 55{,}253 & \$0.36 \\
independent\_ensemble     & 0.159 & $-0.007$ & 0.011 & 0.020 & 0.110 & 15{,}362 & \$0.10 \\
orchestrator\_specialist  & 0.162 & $-0.009$ & 0.011 & 0.025 & 0.112 & 49{,}215 & \$0.31 \\
peer\_critique\_debate    & 0.170 & $-0.017$ & 0.012 & 0.020 & 0.100 & 36{,}015 & \$0.23 \\
consensus\_alignment      & 0.181 & $-0.028$ & 0.015 & 0.026 & 0.094 & 15{,}113 & \$0.10 \\
\midrule
\textit{market consensus} & 0.152 & --- & --- & --- & --- & --- & --- \\
\bottomrule
\end{tabular}
\caption{Empirical leaderboard, claude-opus-4-6, $n=100$, web search disabled. Uncertainty term in the Murphy decomposition $\UNC = 0.249$ for all configurations (a property of the question set, not the forecaster). All five configurations underperform the market-consensus baseline (negative Alpha), with consensus\_alignment showing the largest underperformance.}
\label{tab:leaderboard}
\end{table}

The aggregate ranking is sequential\_pipeline $<$ independent\_ensemble $\approx$ orchestrator\_specialist $<$ peer\_critique\_debate $<$ consensus\_alignment, with the spread between best and worst at 0.028 Brier units. The market-consensus baseline (0.152) is matched by sequential\_pipeline and exceeded (worse) by all other configurations. \emph{No configuration produces positive Alpha on this fixture}---the model with the available tools and web search disabled does not extract edge over the Polymarket consensus on post-cutoff markets at this scale. This is a substantive finding rather than a null result; we discuss the implications in Section~\ref{sec:discussion}.

\subsection{Murphy-decomposition signatures}
\label{sec:results-murphy}

\begin{figure}[h]
\centering
\includegraphics[width=0.85\textwidth]{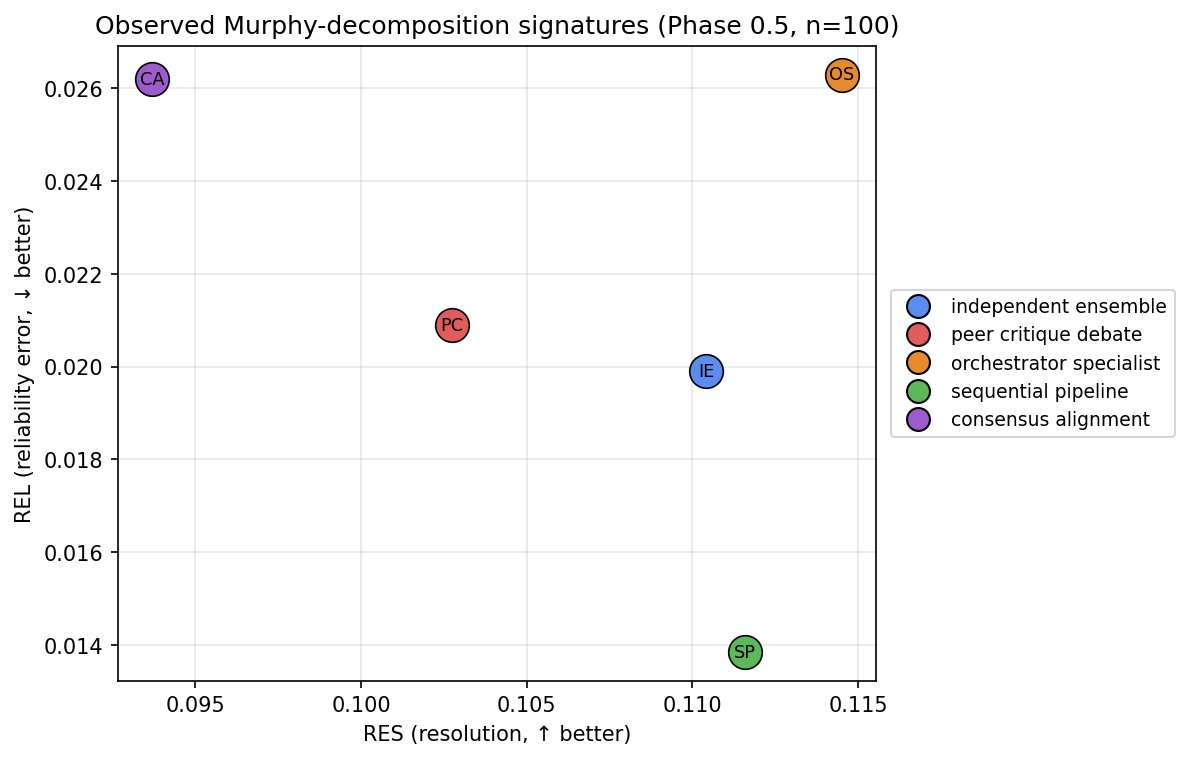}
\caption{Observed Murphy-decomposition signatures of the five configurations on the Phase 0.5 fixture ($n=100$), with the market-consensus baseline implicit (Brier 0.152, decomposing into the model's own REL/RES values). Lower $\REL$ is better calibration; higher $\RES$ is better discrimination. Point positions are reported without confidence regions: at $n=100$ with $K=10$ probability-decile bins, per-bin sample sizes ($\sim 10$ markets) are too small to support stable bin-conditional reliability estimates, and the figure is best read as a diagnostic visualization of relative position rather than as a precise localization. The bin-count sensitivity of these positions (recomputing at $K=5$ and $K=20$, and using equal-mass quantile bins instead of fixed deciles) is reported in the released analysis script.}
\label{fig:observed}
\end{figure}

Figure~\ref{fig:observed} shows the observed positions of the five configurations in REL--RES space, alongside the market-consensus baseline.

The five configurations separate visibly along both axes, but we caution against over-reading exact distances in this figure. With $n=100$ and $K=10$ probability-decile bins, the per-bin counts are small and the resulting REL and RES values carry sampling noise that the figure does not visualize. We make no formal pairwise inference on REL or RES alone here; the pairwise inference in this paper is on Brier (Section~\ref{sec:results-stats} and Section~\ref{sec:results-power}). The figure should be read as showing that configurations occupy \emph{distinguishable regions} of REL--RES space, not that any particular point is precisely located. Comparing the qualitative observed regions to the pre-specified predictions of Figure~\ref{fig:predictions}:

\begin{itemize}
\item \emph{Independent ensemble} (Prediction 1: moderate REL, high RES): \textbf{upheld}. Observed REL = 0.020, observed RES = 0.110---both moderate, with RES at the higher end of the configuration range.
\item \emph{Peer-critique debate} (Prediction 2: REL improves, RES declines): \textbf{partially upheld}. Observed RES (0.100) is the second-lowest among configurations, consistent with diversity-collapse from alignment pressure. Observed REL (0.020) is mid-range rather than the lowest as predicted; cross-correction reduces calibration error somewhat but not as much as the prediction implied.
\item \emph{Orchestrator-specialist} (Prediction 3: low REL, moderate RES): \textbf{not upheld}. Observed REL is the highest among configurations (0.025), not the lowest. The orchestrator imposes a final calibration step but the integrator's uncertainty about specialist outputs appears to add miscalibration rather than reduce it.
\item \emph{Sequential pipeline} (Prediction 4: REL and RES both critically dependent on stage 1): \textbf{partially upheld in best case}. Observed signature is the lowest REL (0.013) and the highest RES (0.109)---which corresponds to the prediction's best-case branch (competent stage 1). On this fixture the researcher stage performed adequately, allowing the analyst and forecaster to elaborate on a useful frame rather than a wrong one.
\item \emph{Consensus alignment} (Prediction 5: very low REL relative to convergence point, very low RES): \textbf{upheld}. Observed RES is the lowest among configurations (0.094); REL is high (0.026), consistent with convergence on a midpoint that misses the eventual outcome distribution. The Foresight Arena empirical observation about market-tracking failure modes reproduces qualitatively here on a single model.
\end{itemize}

Three predictions out of five are upheld in spirit; one (orchestrator-specialist) is contradicted in direction; one (sequential pipeline) corresponds to the prediction's best-case branch rather than the prediction's median expectation. The mixed result is informative: the layer specification generates predictions that are testable rather than vacuous, and the empirical phase has both confirmed and disconfirmed specific claims.

\subsection{Pairwise statistical separation}
\label{sec:results-stats}

We compute paired t-tests on per-market squared errors across all $\binom{5}{2} = 10$ configuration pairs, using the 94 markets common to every configuration's successful prediction set. Smallest p-values:

\begin{itemize}
\item consensus\_alignment vs orchestrator\_specialist: $p = 0.075$
\item consensus\_alignment vs sequential\_pipeline: $p = 0.080$
\item consensus\_alignment vs independent\_ensemble: $p = 0.082$
\end{itemize}

No pair reaches $p < 0.05$ uncorrected, and the Bonferroni-corrected threshold across 10 pairs is $p < 0.005$. At $n=100$ the observed Brier-spread of 0.028 is approximately $2 \times$ the per-configuration SEM, which is suggestive but not statistically resolved. Power analysis from \citet{nechepurenko2026foresight} indicates that detecting a true Alpha difference of 0.02 at $\kappa=0.05$, $\pi=0.80$ requires approximately 350 paired observations; resolving the smaller adjacent gaps observed here (0.006--0.011) would require closer to 1{,}500--2{,}000.

\subsection{Cost--quality Pareto frontier}
\label{sec:results-pareto}

\begin{figure}[h]
\centering
\includegraphics[width=0.85\textwidth]{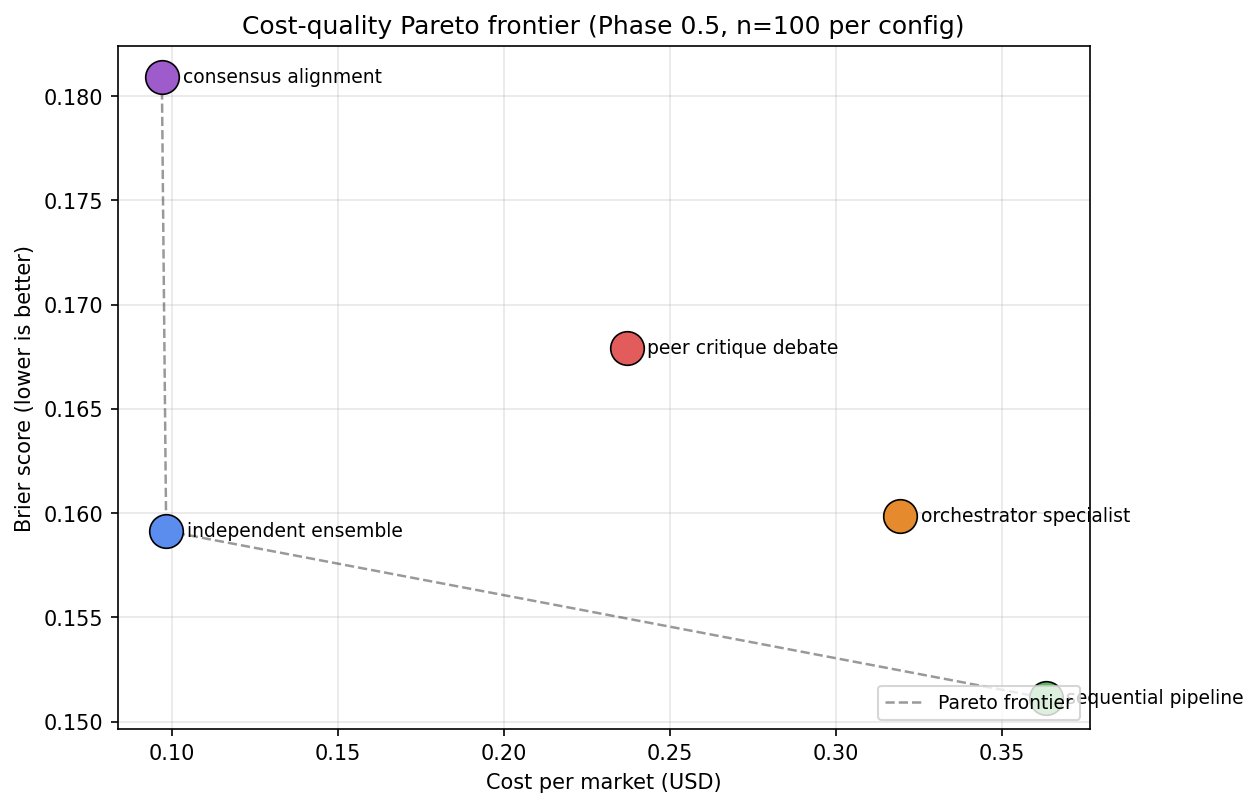}
\caption{Cost--quality Pareto frontier on Phase 0.5 ($n=100$ per configuration). The dashed line connects non-dominated configurations. Independent ensemble dominates orchestrator-specialist and peer-critique-debate on cost-adjusted accuracy; sequential pipeline achieves the lowest Brier at the highest cost.}
\label{fig:pareto}
\end{figure}

Figure~\ref{fig:pareto} plots per-market cost against Brier score.

Two configurations form the frontier: \texttt{independent\_ensemble} at \$0.10/market (Brier 0.159) and \texttt{sequential\_pipeline} at \$0.36/market (Brier 0.153). \texttt{Orchestrator\_specialist} (\$0.31, Brier 0.162) and \texttt{peer\_critique\_debate} (\$0.23, Brier 0.170) are both dominated: each costs more than ensemble while delivering worse Brier. \texttt{Consensus\_alignment} (\$0.10, Brier 0.181) is dominated by ensemble at the same cost point.

This is a separable empirical finding regardless of statistical resolution: even if the Brier differences in Table~\ref{tab:leaderboard} ultimately fail Bonferroni at any sample size, the cost differences are large and reproducible. Within this implementation, this information regime, this model, and this task family, orchestrator-specialist and peer-critique-debate are dominated by independent ensemble on the observed cost--quality frontier. We caveat the generalization explicitly: the relative cost of each configuration is partly a property of architecture (how many calls, how much context propagates) and partly a property of implementation (whether the harness re-sends or caches the system prompt, how aggressively tool result payloads are summarized). A more cache-aware harness could lower orchestrator's cost without changing its architecture; a model with cheaper input tokens would compress all cost differences. The released harness allows external researchers to run this same comparison under different implementation choices and we encourage that.

\subsection{Category-conditional analysis}
\label{sec:results-categories}

Table~\ref{tab:per-category} reports per-configuration Brier broken down by category.

\begin{table}[h]
\centering
\small
\begin{tabular}{lcccccc|c}
\toprule
Configuration & crypto & economics & entertainment & geopolitics & politics & sports & overall \\
\midrule
sequential\_pipeline      & 0.080 & 0.120 & 0.169 & 0.201 & 0.158 & 0.178 & 0.151 \\
independent\_ensemble     & 0.100 & 0.196 & 0.169 & 0.186 & 0.140 & 0.166 & 0.159 \\
orchestrator\_specialist  & 0.081 & 0.205 & 0.177 & 0.190 & 0.145 & 0.168 & 0.160 \\
peer\_critique\_debate    & 0.086 & 0.241 & 0.170 & 0.212 & 0.143 & 0.166 & 0.168 \\
consensus\_alignment      & 0.095 & 0.231 & 0.214 & 0.240 & 0.144 & 0.166 & 0.181 \\
\bottomrule
\end{tabular}
\caption{Per-category Brier scores. Within-row spread varies dramatically by category: economics shows a 0.121 spread (sequential 0.120 vs peer-critique 0.241), while sports shows a 0.012 spread. The ``overall'' column aggregates over only the successful predictions per configuration and so differs slightly from Table~\ref{tab:leaderboard} for configurations with non-zero failure counts (e.g.\ sequential\_pipeline 0.151 here vs 0.153 in Table~\ref{tab:leaderboard}, reflecting the two markets that fell to the fallback in Section~\ref{sec:results-leaderboard}).}
\label{tab:per-category}
\end{table}

The architectural effect is category-conditional. Sports markets show essentially no separation between configurations (0.012 spread); economics markets show a $\sim 6\times$ larger spread (0.121). Crypto and politics fall in the middle. The sequential pipeline configuration is best in crypto, economics, and worst in geopolitics; consensus\_alignment is uniformly worst except in sports where all configurations roughly tie.

This pattern is consistent with a hypothesis we did not pre-register but is suggested by the data: coordination architecture matters more in domains where the question requires structured numerical reasoning (economics) than in domains where it does not (sports). The pattern is too underpowered to claim formally on $n \approx 17$ per category, and we flag it as a candidate hypothesis for the follow-up.

\subsection{Power projection from observed data}
\label{sec:results-power}

The pairwise statistical separation in Section~\ref{sec:results-stats} did not reach Bonferroni-corrected significance at $n=100$. A natural next question is: how much data would be needed to resolve which pairs, and how confident should we be in the direction and magnitude of the observed effects? Rather than scaling the experiment by an arbitrary factor and re-running, we extract this information from the existing sample using three standard procedures from observational power analysis: paired bootstrap of the per-pair Brier difference; required-$n$ projection under the observed effect size; and Type-S / Type-M error analysis~\citep{cohen1988statistical,gelman2014beyond}. The resampling unit throughout is the per-market squared error, the natural unit for paired comparisons in this design.

\begin{figure}[h]
\centering
\includegraphics[width=0.95\textwidth]{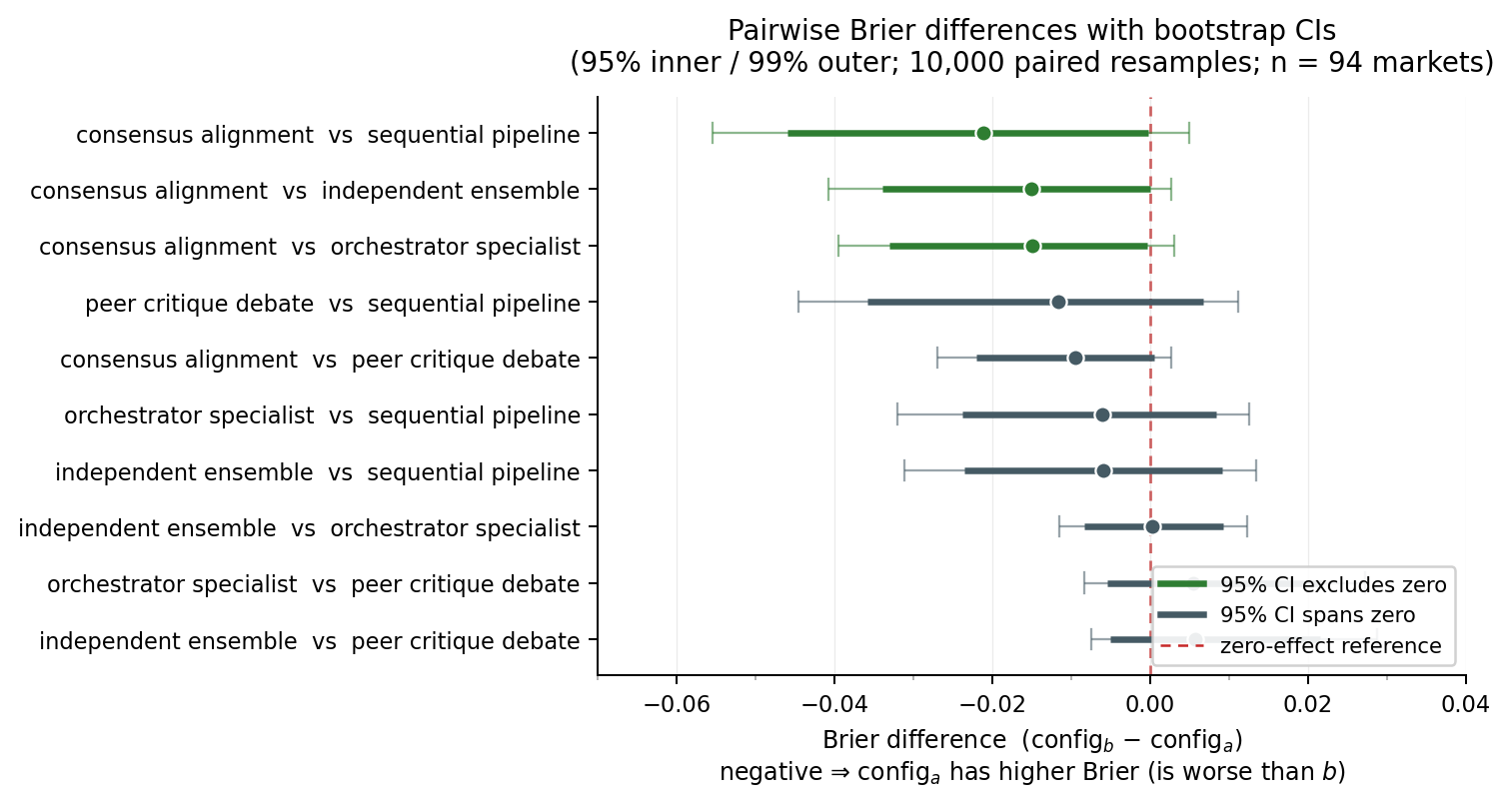}
\caption{Pairwise Brier differences with bootstrap CIs. The 95\% inner intervals (thick bars) and 99\% outer intervals (thin bars) are computed from 10{,}000 paired resamples on $n=94$ common scored markets. Negative values indicate that \texttt{config\_a} has higher Brier (is worse). Three pairs involving \texttt{consensus\_alignment}---versus sequential\_pipeline, independent\_ensemble, and orchestrator\_specialist respectively---have 95\% CIs that exclude zero (green), providing the strongest evidence of architectural separation in the dataset.}
\label{fig:bootstrap}
\end{figure}

\paragraph{Bootstrap pairwise differences.}
For each of the 10 configuration pairs, we resample the 94 paired squared errors with replacement (10{,}000 resamples) and recompute the mean Brier difference. Figure~\ref{fig:bootstrap} shows the resulting forest plot with 95\% (inner) and 99\% (outer) confidence intervals.

The forest plot reveals what the t-test summary of Section~\ref{sec:results-stats} compresses: three pairs (\texttt{consensus\_alignment} vs \texttt{sequential\_pipeline}, vs \texttt{independent\_ensemble}, vs \texttt{orchestrator\_specialist}) have 95\% bootstrap CIs that exclude zero (bootstrap probabilities $p$ that consensus is better: 0.023, 0.024, 0.020 respectively). The remaining seven pairs have CIs that straddle zero. The architectural separation is concentrated: it is consensus alignment versus the other configurations that is most strongly resolved on this sample, while contrasts not involving consensus alignment remain ambiguous at $n=94$.

\paragraph{Reconciling bootstrap and t-test results.}
The bootstrap finding (three 95\% CIs exclude zero) and the paired-t-test finding ($p \approx 0.075$--$0.082$ for the same three pairs) are computed on the same paired sample of $n = 94$ common scored markets. Both are two-sided. The CIs are percentile bootstrap intervals constructed from 10{,}000 resamples; the t-test p-values use the standard paired-difference formula. The two methods disagree on the borderline because they make different distributional assumptions: the t-test assumes the per-market squared-error differences are approximately normal (which they are not---the distribution is heavy-tailed because some markets resolve into the corner of the prediction space), while the percentile bootstrap is distribution-free but borrows strength from the asymmetry of the empirical distribution. For the three consensus-alignment pairs, the empirical distribution of paired differences is asymmetric in the direction that makes the bootstrap reject zero slightly more readily than the t-test. Neither method is wrong; they answer slightly different questions. We report both and treat the bootstrap CIs as \emph{exploratory} indicators of separation rather than confirmatory significance claims, especially in light of the multiple-comparisons issue addressed below.

\paragraph{Multiple comparisons and exploratory framing.}
With 10 pairwise comparisons, the Bonferroni-corrected threshold for the t-tests is $p < 0.005$, which is not reached by any pair. Applying the same correction philosophy to the bootstrap intervals (i.e., requiring 99.5\% rather than 95\% CIs) yields 99\% CIs that all straddle zero (the outer thin bars in Figure~\ref{fig:bootstrap}). The bootstrap finding therefore should be read as: \emph{exploratory evidence that consensus alignment separates from the other configurations more strongly than any other pair on this sample}, not as a corrected pairwise significance result. The required-$n$ analysis below quantifies the follow-up sample sizes needed to resolve these contrasts under proper correction.

\begin{figure}[h]
\centering
\includegraphics[width=0.95\textwidth]{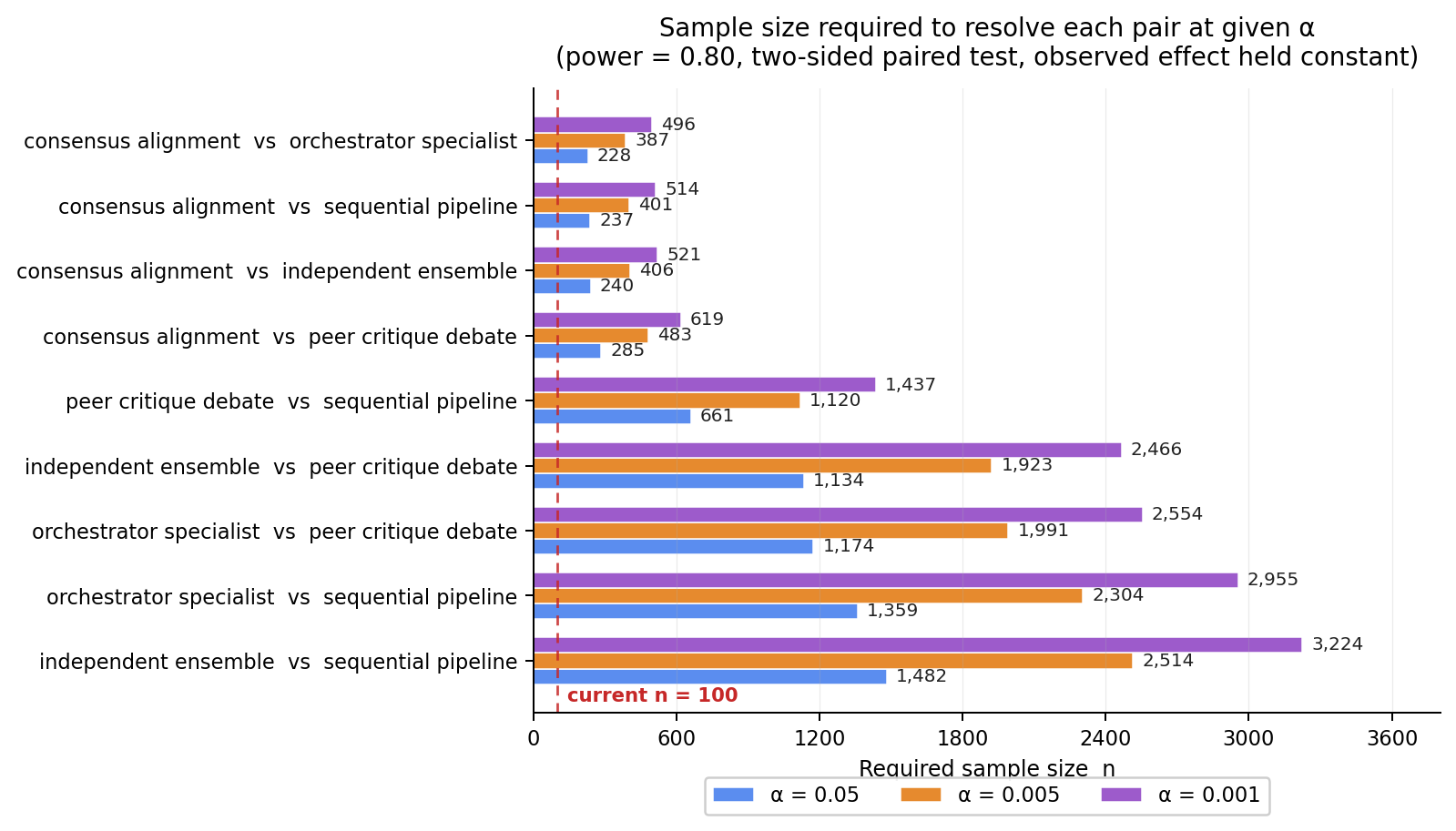}
\caption{Sample size required to resolve each pair at three significance thresholds, holding the observed effect size constant (power 0.80, two-sided paired test). Pairs are sorted by required-$n$ at $\alpha = 0.005$ (the Bonferroni-corrected threshold for 10 pairs). The current sample size $n = 100$ is marked. The pair \texttt{independent\_ensemble} vs \texttt{orchestrator\_specialist} is omitted from the plot: its observed Brier difference is $+0.0002$, indistinguishable from zero on this sample, and the corresponding required $n$ exceeds one million at every $\alpha$ level. We treat that pair as not having a meaningful detectable effect at this experimental setup.}
\label{fig:required-n}
\end{figure}

\paragraph{Required sample size.}
For each pair, we compute the sample size required to detect the observed effect at three significance levels ($\alpha = 0.05$ uncorrected; $\alpha = 0.005$ Bonferroni-corrected for 10 pairs; $\alpha = 0.001$ stringent), with power $0.80$ in a two-sided paired test. Results in Figure~\ref{fig:required-n}.

The pairs separate into three regimes:
\begin{itemize}
\item \textbf{Resolvable at $n \in [387, 483]$}: the four \texttt{consensus\_alignment} pairs. A follow-up at $n \approx 500$ would resolve all four at the Bonferroni-corrected threshold $\alpha = 0.005$ with power 0.80.
\item \textbf{Resolvable at $n \in [1{,}120, 2{,}514]$}: \texttt{peer\_critique\_debate} vs \texttt{sequential\_pipeline} (1{,}120), and the four pairs that compare \texttt{independent\_ensemble} or \texttt{orchestrator\_specialist} against \texttt{peer\_critique\_debate} or \texttt{sequential\_pipeline} (1{,}923 to 2{,}514). A follow-up at $n \approx 2{,}500$ would resolve all of them.
\item \textbf{Indistinguishable on this sample}: \texttt{independent\_ensemble} vs \texttt{orchestrator\_specialist}, with observed difference $+0.0002$ and required $n$ exceeding $10^6$. Whether the true underlying effect is exactly zero or merely smaller than the experiment can detect, the substantive interpretation is the same: with the current information regime, prompts, and tool stack, these two configurations are not empirically separable.
\end{itemize}

This ranking is the actionable output of the power analysis: it identifies which architectural contrasts the data already supports, which can be confirmed at moderate additional cost, and which the data suggests are not worth pursuing further at this experimental setup.

\paragraph{Type-S and Type-M errors.}
For each pair, we compute the probabilities of sign error (Type-S) and magnitude inflation (Type-M) under the assumption that the true effect equals the observed effect, following \citet{gelman2014beyond}. For the four \texttt{consensus\_alignment} pairs, Type-S errors are below 0.5\% and Type-M is in the range 1.49--1.63: the sign of the effect is well established and the magnitude is at most modestly inflated. For the seven other pairs, Type-S rises to 2--4\% (still below conventional detection thresholds) and Type-M to 2.4--3.5, indicating that any future single-instance ``confirmation'' on these noisier pairs would likely overstate the magnitude even if the sign were correct, and we caution against design choices motivated by such future confirmations.

\paragraph{Implication for follow-up scope.}
The combined output of bootstrap, required-$n$, and Type-S/M analysis re-frames the follow-up question. The naive question is ``how many markets do we need to resolve all 10 pairs?''---a question whose answer is dominated by the consensus-of-noise pair (\texttt{independent\_ensemble} vs \texttt{orchestrator\_specialist}) and is therefore unbounded. The empirically grounded question is ``which pairs are worth resolving, and at what cost?''---answered by the three-tier classification above. A follow-up at $n \approx 500$ would close out the consensus-alignment claims; an extension to $n \approx 2{,}500$ would resolve the four pairs that contrast the Pareto-frontier configurations against the dominated configurations. Beyond that, the marginal information yield per market drops sharply, and the experimental budget is more productively spent on a different information regime (Section~\ref{sec:discussion-info-env}, Section~\ref{sec:results-replication}) than on chasing remaining within-regime contrasts.

\subsection{Production validation channel: Foresight Arena}
\label{sec:results-replication}

In parallel with the historical-sandbox study reported above, we deploy the same five coordination configurations as live agents on Foresight Arena~\citep{nechepurenko2026foresight}, an open on-chain forecasting benchmark on Polymarket. The deployment uses identical role prompts, identical aggregation logic, identical Murphy-decomposition scoring---and crucially, web search \emph{enabled} on real future events with short resolution windows. Agents are registered under the prefix \texttt{foreflow-*} (for example \texttt{foreflow-ensemble}, \texttt{foreflow-pipeline}, \texttt{foreflow-debate}, \texttt{foreflow-orchestrator}, \texttt{foreflow-consensus}) and submit predictions on Polymarket via the ERC-8004 commit-reveal protocol described in the companion paper.

This channel exists for three purposes. First, methodological replication under a deliberately different information regime: the historical sandbox controlled information access by disabling web search to eliminate inter-call variation; the production channel controls it instead by exposing the same web-search tool to every configuration. The architectural specification is unchanged, but the Information Fixing principle (Section~\ref{sec:method-principle}) acquires a new variant that this paper does not yet test. Second, independent on-chain attestation: predictions are committed before resolution and revealed after, providing a tamper-evident record of when a probability was assigned and what the model knew at that moment. Third, public visibility: live results are observable on the Foresight Arena dashboard as predictions accumulate, and any reader can verify the ongoing replication by examining the on-chain record.

The rate of data accumulation is set by the Foresight Arena round cadence (1--2 rounds per day in the current deployment), so a full power-equivalent replication of the historical sandbox is expected to take several months. This separation of timescales is by design: the present paper is a methodology-validating study with sufficient sample size for the contrasts that its observational power analysis identifies as resolvable, while the production channel accumulates the data for the information-regime extension. We report on the production-channel results in a follow-up paper as data becomes available; the deployment infrastructure and agent code are released as a public artifact (Section~\ref{sec:experiment-artifacts}).

\subsection{Failure-mode case studies}
\label{sec:results-cases}

Three markets in the fixture exhibit the largest cross-configuration probability disagreement and serve as qualitative case studies for failure modes implicit in the layer specification.

\paragraph{Minority-view collapse in peer-critique debate.}
On market 66 (\textit{Will WTI Crude Oil hit \$100 LOW in April?}, baseline 0.587, outcome YES), sequential\_pipeline assigned $p = 0.85$, independent\_ensemble $p = 0.58$, orchestrator\_specialist $p = 0.55$, while peer\_critique\_debate dropped to $p = 0.12$ and consensus\_alignment to $p = 0.15$. The two configurations involving multi-round inter-agent communication converged on a strong NO position against three other configurations holding moderate-to-high YES. The pre-specified Prediction~2 anticipated alignment-pressure suppression of minority dissent; this market is a concrete instance, observed without us having selected for it.

\paragraph{Midpoint anchoring in consensus alignment.}
On market 92 (\textit{Renate Reinsve Best Actress nomination}, baseline 0.873, outcome YES), four configurations assigned $p \in [0.83, 0.87]$ (matching the market consensus); consensus\_alignment converged at $p = 0.35$. The other agents in consensus\_alignment did not all start at 0.35---they converged there from initially diverse positions. This is the convergence-to-midpoint failure mode anticipated by Prediction~5.

\paragraph{False confidence cascade in sequential pipeline.}
On market 32 (\textit{Spread: Bills (-7.5)}, baseline 0.502, outcome YES), four configurations all output $p = 0.50$ (refusing to forecast in the absence of information); sequential\_pipeline output $p = 0.25$. The pipeline's structure prevents stage-3 from refusing to commit if stages 1 and 2 produced any analysis at all---even when that analysis is a polite gloss on no real evidence. This is the cascade failure mode anticipated by Prediction~4 in its non-best-case branch, and it is consistent with the observation that sequential\_pipeline's category-best wins (crypto, economics) come paired with category-worst losses (geopolitics) at higher confidence than the other configurations.

% =====================================================================
\section{Discussion}
\label{sec:discussion}

\subsection{What the methodology validates}
\label{sec:discussion-validates}

We frame the contribution narrowly. This paper does not establish a general law that coordination-layer specifications predict failure-mode signatures across models, domains, or information regimes. It establishes one weaker but still useful claim: that coordination-layer specifications can \emph{generate} falsifiable failure-signature hypotheses in advance of running an experiment, and that on a single first instantiation those hypotheses are not vacuous---some are upheld, some are contradicted, some are conditionally upheld in ways predicted by the specification itself. This is the difference between a diagnostic methodology that produces testable predictions and a predictive theory that closes them. We are claiming the former.

The information-controlled design produces visible architectural separation along Murphy axes that would be invisible to single-Brier comparison. Table~\ref{tab:leaderboard} alone shows configurations whose Brier scores differ by less than the per-configuration SEM, but Figure~\ref{fig:observed} shows them occupying distinct regions of REL--RES space. The decomposition is doing the work it is supposed to.

The pre-specified predictions are testable rather than vacuous: three are upheld (Predictions 1, 2 partially, 5), one is contradicted (Prediction 3), one matches its conditional best-case branch (Prediction 4). A framework that produced 5/5 confirmation on a first instantiation would be more concerning than this mixed pattern, because the predictions would be too coarse to fail.

The qualitative failure modes anticipated by the layer specification reproduce on individual markets in the fixture (Section~\ref{sec:results-cases}) without our having selected for them. The connection between architectural choice and concrete behavioral failure---the connection that motivated this paper's central claim---is observable in this small dataset.

\subsection{What the methodology does not yet validate}
\label{sec:discussion-not-yet}

Statistical power at $n=100$ is insufficient to resolve adjacent configuration pairs. Three pairs (consensus\_alignment vs orchestrator\_specialist / sequential\_pipeline / independent\_ensemble) reach $p \approx 0.075$--$0.082$, suggestive but not Bonferroni-significant. The observed Brier spread of 0.028 is consistent with the predicted effect size $0.02$--$0.05$ from Section~\ref{sec:layer-configs}, and a follow-up at $n \approx 500$--$700$ would be expected to resolve at least the consensus-alignment-vs-others contrast at corrected significance, with $n \approx 1{,}500$--$2{,}000$ required to resolve all pairs.

The empirical scope is one model, one task family, one tool stack with web search disabled. Cross-model robustness, alternative tool stacks, and domain transfer are explicitly out of scope.

\subsection{The information-environment finding}
\label{sec:discussion-info-env}

A finding we did not anticipate at the methodology stage: \emph{no configuration achieved positive Alpha against the Polymarket consensus} (Table~\ref{tab:leaderboard}). The market-consensus baseline matches the best configuration (sequential pipeline) and exceeds the rest. This is not a surprise once stated, but its implication for the methodology is worth articulating.

The methodology of Section~\ref{sec:method} fixed information access uniformly across configurations. We achieved this by, among other things, disabling web search. This is methodologically clean: it eliminates a potentially massive confound, since web search results differ between requests in ways that can dwarf coordination effects. But it also creates a regime in which the agent's only inputs about a market are the model's training data plus the structured Polymarket metadata. For markets that resolve after the model's training cutoff and depend on current-events knowledge, the agent has, by construction, less information than the human-and-bot mixture trading the market in real time. The agent is being asked to forecast events partially in its informational future, against participants in its informational present.

The negative-Alpha-everywhere result is therefore not a verdict on coordination architectures. It is a verdict on what the architectures we studied can do given the informational constraint we imposed. This separates two distinct empirical questions which the methodology, as framed, can answer:

\begin{enumerate}
\item Given fixed information access, which coordination configurations preserve / damage / improve the model's ability to extract signal from that information? \emph{This is what Phase 0.5 measures.}
\item Which coordination configurations make better use of access to richer information (web search, deep retrieval, real-time tool use)? \emph{This is a distinct question requiring web search ON or equivalent retrieval, ideally on real future events.}
\end{enumerate}

The current paper measures (1). The framework specification is unchanged for (2)---it is the same five configurations, the same Murphy decomposition, the same propriety guarantees---but the methodology requires updating to fix the new variant of information access uniformly. Web search results, unlike the static Polymarket metadata, vary between requests, and the methodology must extend to control for that variation.

We identify (2) as the natural follow-up that future work should address. Concretely, the live Foresight Arena deployment described in Section~\ref{sec:results-replication} is the experimental channel for that follow-up, and predictions from the deployed agents are accumulating as this paper is being written. The two-paper structure---static-information-regime methodology validation here, tool-augmented-information-regime empirical study in the follow-up---is a methodological choice rather than an artifact of resourcing. Mixing the two regimes in a single study would have made the contributions harder to attribute and the threats to validity (especially the prompt-sensitivity-meets-search-result-variation interaction) much harder to control.

\subsection{Implications for declarative-orchestration framework design}
\label{sec:discussion-frameworks}

The Pareto frontier in Figure~\ref{fig:pareto} contains a finding that is independently significant for declarative-orchestration framework users (\citet{daunis2025declarative}, \citet{auton2026}, \citet{strands2025}, \citet{foundry2026}). Of the five reference configurations, two dominate the cost--quality space and three are dominated. A practitioner choosing a coordination pattern from a declarative-framework menu has, on this task family, one cost-sensitive choice (independent ensemble) and one quality-sensitive choice (sequential pipeline)---and three patterns that they should not pick. The dominated patterns are not the patterns one would intuitively expect to be poor (orchestrator-specialist and peer-critique-debate are both popular in the literature and in framework documentation).

This is the kind of finding that the architectural-prediction perspective makes available and that the engineering-productivity perspective on declarative frameworks does not. A framework that markets ``faster deployment'' for orchestrator-specialist as a feature is not wrong about that, but on this task family it is making the deployment of a dominated configuration faster.

\subsection{Mapping observed signatures to MAST failure modes}
\label{sec:discussion-mast}

The MAST taxonomy of \citet{cemri2025mast} identifies three top-level failure clusters: system-design issues, inter-agent misalignment, and task-verification gaps. The Phase 0.5 case studies (Section~\ref{sec:results-cases}) instantiate two of these:

\begin{itemize}
\item Market 66 (peer-critique minority-view collapse) is a clear instance of \emph{communication breakdown} (FM-2.x in MAST), specifically the variant in which alignment pressure suppresses correct minority signals \citep{wynn2025talk}.
\item Market 32 (sequential-pipeline false confidence) is a clear instance of \emph{inadequate output checking} (FM-3.x in MAST): downstream stages did not verify that upstream evidence was substantive before committing a confident probability.
\item Market 92 (consensus-alignment midpoint anchoring) does not map cleanly to a single MAST entry, but is closest to FM-2.4 (state desynchronization where agents converge on different interpretations of the same target).
\end{itemize}

Predicting which of these MAST failure modes a particular configuration is most likely to exhibit \emph{from the configuration's specification alone}---the goal articulated in Section~\ref{sec:related-failures}---appears feasible at the resolution of this experiment. A configuration with all-to-all peer communication and round-bounded termination (peer-critique debate, consensus alignment) is a plausible breeding ground for FM-2; a configuration with strict stage ordering and no architectural opportunity for upstream verification (sequential pipeline) is a plausible breeding ground for FM-3.

\subsection{Threats to validity}
\label{sec:discussion-threats}

\paragraph{Prompt sensitivity.} The role-specific instruction blocks were authored once and not subjected to systematic ablation. A more careful study would vary the role prompts within each configuration while keeping the configuration's structure fixed, to attribute observed effects to coordination structure rather than to particular phrasings.

\paragraph{Sample-size frontier.} The fixture is balanced by category and by baseline decile but contains only $\sim 17$ markets per category. Per-category claims (Section~\ref{sec:results-categories}) are flagged as suggestive rather than tested.

\paragraph{Information leakage via baseline price and price history.} The agents do not see the eventual outcome (which resolves after the model's training cutoff and after the fixture's baseline-price snapshot timestamp), so memorization of outcomes is excluded. However, the agents do see the baseline mid-price and the tool \texttt{getPriceHistory} returns historical CLOB prices up to the baseline timestamp. These prices embed aggregate human-and-bot trader information accumulated after the model's training cutoff but before the resolution event. Consequently, the task this paper measures is not ``forecast a future event from the model's own knowledge.'' It is more accurately: ``produce a probability calibrated against the market's prior consensus, using whatever the model brings beyond what the market already incorporates.'' The negative-Alpha-everywhere result of Section~\ref{sec:discussion-info-env} is consistent with this framing: with web search disabled, agents have nothing additional to bring relative to the market consensus that the price history already encodes. The baseline price serves as the comparator (intentional methodology), not as a hidden label. We treat this not as a confound but as the substantive task definition; the architectural-layer claim is that different coordination configurations process the same market-implied information differently, not that any of them discover information independently of the market.

\paragraph{Outcome adjudication.} Polymarket UMA-arbitrated outcomes were used as ground truth. UMA arbitration has known disputes; for this fixture we excluded markets flagged as disputed.

\paragraph{Single sandbox.} Only the historical-sandbox variant of the methodology was instantiated. The API-sandbox and on-chain production variants would test the same predictions in different information regimes; their absence is acknowledged as a scope limitation, not as a methodological flaw, and addressed in Section~\ref{sec:conclusion}.

% =====================================================================
\section{Conclusion and Future Work}
\label{sec:conclusion}

We argued that coordination in LLM-based multi-agent systems should be treated as a configurable architectural layer separable from agent logic and information access, and that this separation enables falsifiable predictions about failure-mode signatures rather than only engineering productivity. We instantiated the framework with five reference configurations, pre-specified Murphy-decomposition signatures for each, and tested them on 100 post-cutoff Polymarket binary markets under uniform information access. Three of five pre-specified predictions were upheld in the predicted direction; two were contradicted or upheld only conditionally. Within this implementation and information regime, two configurations form the cost--quality Pareto frontier and the other three are dominated. Statistical resolution of adjacent configuration pairs at Bonferroni-corrected significance is not yet achieved at $n=100$; an observational power analysis identifies which contrasts a follow-up at $n \approx 500$ would close out and which require substantially more data. We position this work as a methodology-validating first instantiation of the architectural-layer framework, not as a general claim about cross-model or cross-domain architectural laws.

\paragraph{Released artifacts.}
Three public repositories are released alongside this paper, documented in Section~\ref{sec:experiment-artifacts}: the experimental harness (\href{https://github.com/ForesightFlow/coordination-experiment}{coordination-experiment}), the open trace dataset (\href{https://github.com/ForesightFlow/datasets}{coordination-traces-100}), and the production-agent deployment (\href{https://github.com/ForesightFlow/foreflow-agents}{foreflow-agents}). The version snapshots used in this paper are tagged \texttt{paper-v05} for exact reproduction.

\paragraph{Follow-up work.}

\begin{enumerate}
\item \textbf{Information-regime extension on accumulated production data.} The Foresight Arena deployment (Section~\ref{sec:results-replication}) is accumulating predictions under web-search-enabled conditions on real future events. As the on-chain record reaches the sample sizes that Section~\ref{sec:results-power} identifies as resolvable for each architectural pair, a follow-up paper will report cross-architecture Murphy decomposition under that information regime and contrast it with the static-information regime of this paper. We anticipate that some configurations whose architectural cost is invisible in the static regime will exhibit it under tool-augmented conditions, and that the current paper's negative-Alpha-everywhere result will not generalise.

\item \textbf{Cross-model robustness.} A separate paper will replicate the Phase 0.5 analysis on at least three frontier models (gpt-5.x family, gemini-3.x family) to test whether the observed REL--RES signatures are properties of the coordination layer or of the model. Prediction~5 (consensus-alignment market-tracking) carried the strongest cross-model claim and would be the principal target.

\item \textbf{Selective resolution at moderate scale.} The bootstrap and required-$n$ analysis of Section~\ref{sec:results-power} identifies four pairs that a follow-up at $n \approx 500$--$700$ markets, in the same information regime, would resolve at Bonferroni-corrected significance. We treat this as a lower-priority extension than the information-regime work, because the architectural claim it would resolve is the within-regime contrast already characterised qualitatively here, while the information-regime extension addresses a distinct empirical question that the current methodology cannot answer.

\item \textbf{ILS-stratified analysis.} ForesightFlow's Information Leakage Score, when production-grade, will allow re-running the same analysis on subsets stratified by likelihood of insider-driven price discovery. We expect architectural signatures to be robust to ILS thresholding but flag this as an empirical question for follow-up.

\item \textbf{Hybrid and adaptive configurations.} The five reference configurations are pre-specified as a fixed grid. Combinations (e.g., orchestrator-dispatched ensembles) and adaptive policies that select coordination structure based on the question's category or difficulty are out of scope here but plausible candidates for the next iteration. The release of the harness as a reusable substrate (Section~\ref{sec:experiment-artifacts}) explicitly enables external researchers to test such configurations against the architectural specification developed here.
\end{enumerate}

The contribution of this paper is the framing, the predictions, the methodology, the first instantiation of that methodology under uniformly fixed information access, and the trio of public artifacts that enables independent extension of any of these. The contribution of the follow-up will be the information-regime extension via the production channel and, separately, the cross-model robustness check. We believe this division is methodologically appropriate: a paper that fixes the framing and the predictions before instantiation is harder to misread than one that introduces them after the fact.

% =====================================================================
\section*{Generative AI Disclosure}
\label{sec:ai-disclosure}
\addcontentsline{toc}{section}{Generative AI Disclosure}

In preparing this manuscript, the authors used Anthropic's Claude Opus 4.7 for copy-editing, for literature search and synthesis across the multi-agent and forecasting evaluation literatures, and for rendering figures from numerical and structural specifications. Claude Code (also produced by Anthropic) was used for implementing the experimental harness described in Section~\ref{sec:experiment} under direct human supervision and review of all code. The empirical results reported in Sections~\ref{sec:results}--\ref{sec:discussion} were generated by running this harness with the underlying model claude-opus-4-6 (Section~\ref{sec:experiment-model}); the underlying model is the experimental subject, not a co-author. All methodology, analysis, predictions, and conclusions are the authors' own; the authors reviewed and edited all AI-generated content and take full responsibility for the final manuscript.

% =====================================================================
\bibliographystyle{plainnat}
\bibliography{refs}

@inproceedings{cemri2025mast,
  author    = {Cemri, Mert and Pan, Melissa Z. and Yang, Shuyi and others},
  title     = {Why Do Multi-Agent {LLM} Systems Fail?},
  booktitle = {Proceedings of the 39th Conference on Neural Information Processing Systems (NeurIPS), Datasets and Benchmarks Track},
  year      = {2025},
  note      = {arXiv:2503.13657},
  url       = {https://arxiv.org/abs/2503.13657}
}

@misc{sid2026,
  author    = {Acharya, Vivek},
  title     = {Semantic Consensus: Process-Aware Conflict Detection and Resolution for Enterprise Multi-Agent {LLM} Systems},
  year      = {2026},
  note      = {arXiv:2604.16339},
  url       = {https://arxiv.org/abs/2604.16339}
}

@misc{wynn2025talk,
  author    = {Wynn, Andrea and Satija, Harsh and Hadfield, Gillian},
  title     = {Talk Isn't Always Cheap: Understanding Failure Modes in Multi-Agent Debate},
  year      = {2025},
  note      = {arXiv:2509.05396}
}

@misc{xu2026rethinking,
  author    = {Xu, Zhiyu and Liu, Yanqi and Peng, Hao and others},
  title     = {Rethinking the Value of Multi-Agent Workflow: A Strong Single-Agent Baseline},
  year      = {2026},
  note      = {arXiv:2601.12307}
}

@misc{xia2024agentless,
  author    = {Xia, Chunqiu Steven and Deng, Yinlin and Dunn, Soren and Zhang, Lingming},
  title     = {Agentless: Demystifying {LLM}-Based Software Engineering Agents},
  year      = {2024},
  booktitle = {ACM SIGSOFT International Symposium on Foundations of Software Engineering}
}

@misc{reliability2026,
  author    = {Ao, Ruicheng and Gao, Siyang and Simchi-Levi, David},
  title     = {On the Reliability Limits of {LLM}-Based Multi-Agent Planning},
  year      = {2026},
  note      = {arXiv:2603.26993},
  url       = {https://arxiv.org/abs/2603.26993}
}

@misc{daunis2025declarative,
  author    = {Daunis, Ivan and others},
  title     = {A Declarative Language for Building and Orchestrating {LLM}-Powered Agent Workflows},
  year      = {2025},
  note      = {arXiv:2512.19769}
}

@misc{auton2026,
  author    = {Cao, Sheng and Chang, Zhao and Li, Chang and Li, Hannan and Fu, Liyao and Tang, Ji},
  title     = {The {Auton} Agentic {AI} Framework: A Declarative Architecture for Specification, Governance, and Runtime Execution of Autonomous Agent Systems},
  year      = {2026},
  note      = {arXiv:2602.23720},
  url       = {https://arxiv.org/abs/2602.23720}
}

@misc{strands2025,
  author    = {Hu, Jia and others},
  title     = {Multi-Agent Collaboration Patterns with {Strands} Agents and {Amazon} {Nova}},
  year      = {2025},
  howpublished = {AWS Machine Learning Blog},
  url       = {https://aws.amazon.com/blogs/machine-learning/multi-agent-collaboration-patterns-with-strands-agents-and-amazon-nova/}
}

@misc{foundry2026,
  author    = {{Microsoft}},
  title     = {Foundry Agent Service: Workflow Agents and Declarative Orchestration},
  year      = {2026},
  howpublished = {Microsoft Learn Documentation},
  url       = {https://learn.microsoft.com/en-us/azure/foundry/agents/overview}
}

@misc{wu2023autogen,
  author    = {Wu, Qingyun and Bansal, Gagan and Zhang, Jieyu and others},
  title     = {{AutoGen}: Enabling Next-Gen {LLM} Applications via Multi-Agent Conversation},
  year      = {2023},
  note      = {arXiv:2308.08155}
}

@inproceedings{agashe2025llmcoord,
  author    = {Agashe, Saaket and Fan, Yue and Reyna, Anthony and Wang, Xin Eric},
  title     = {{LLM}-Coordination: Evaluating and Analyzing Multi-Agent Coordination Abilities in Large Language Models},
  booktitle = {Findings of the Association for Computational Linguistics: NAACL},
  year      = {2025},
  note      = {arXiv:2310.03903}
}

@book{wooldridge2009introduction,
  author    = {Wooldridge, Michael},
  title     = {An Introduction to MultiAgent Systems},
  edition   = {2},
  publisher = {Wiley},
  year      = {2009}
}

@article{stone2000multiagent,
  author    = {Stone, Peter and Veloso, Manuela},
  title     = {Multiagent Systems: A Survey from a Machine Learning Perspective},
  journal   = {Autonomous Robots},
  volume    = {8},
  number    = {3},
  pages     = {345--383},
  year      = {2000}
}

@book{gamma1994design,
  author    = {Gamma, Erich and Helm, Richard and Johnson, Ralph and Vlissides, John},
  title     = {Design Patterns: Elements of Reusable Object-Oriented Software},
  publisher = {Addison-Wesley},
  year      = {1994}
}

@article{gelernter1985linda,
  author    = {Gelernter, David},
  title     = {Generative Communication in {Linda}},
  journal   = {ACM Transactions on Programming Languages and Systems},
  volume    = {7},
  number    = {1},
  pages     = {80--112},
  year      = {1985}
}

@book{milner1999pi,
  author    = {Milner, Robin},
  title     = {Communicating and Mobile Systems: The {$\pi$}-calculus},
  publisher = {Cambridge University Press},
  year      = {1999}
}

@book{hoare1985csp,
  author    = {Hoare, C. A. R.},
  title     = {Communicating Sequential Processes},
  publisher = {Prentice Hall},
  year      = {1985}
}

@article{hewitt1973universal,
  author    = {Hewitt, Carl and Bishop, Peter and Steiger, Richard},
  title     = {A Universal Modular {ACTOR} Formalism for Artificial Intelligence},
  journal   = {Proceedings of the 3rd International Joint Conference on Artificial Intelligence},
  pages     = {235--245},
  year      = {1973}
}

@article{brier1950verification,
  author    = {Brier, Glenn W.},
  title     = {Verification of Forecasts Expressed in Terms of Probability},
  journal   = {Monthly Weather Review},
  volume    = {78},
  number    = {1},
  pages     = {1--3},
  year      = {1950}
}

@article{murphy1973decomposition,
  author    = {Murphy, Allan H.},
  title     = {A New Vector Partition of the Probability Score},
  journal   = {Journal of Applied Meteorology},
  volume    = {12},
  number    = {4},
  pages     = {595--600},
  year      = {1973}
}

@article{gneiting2007strictly,
  author    = {Gneiting, Tilmann and Raftery, Adrian E.},
  title     = {Strictly Proper Scoring Rules, Prediction, and Estimation},
  journal   = {Journal of the American Statistical Association},
  volume    = {102},
  number    = {477},
  pages     = {359--378},
  year      = {2007}
}

@article{degroot1983comparison,
  author    = {DeGroot, Morris H. and Fienberg, Stephen E.},
  title     = {The Comparison and Evaluation of Forecasters},
  journal   = {The Statistician},
  volume    = {32},
  number    = {1/2},
  pages     = {12--22},
  year      = {1983}
}

@misc{halawi2024approaching,
  author    = {Halawi, Danny and Zhang, Fred and Yueh-Han, Cheng and Steinhardt, Jacob},
  title     = {Approaching Human-Level Forecasting with Language Models},
  year      = {2024},
  note      = {arXiv:2402.18563}
}

@article{schoenegger2024silicon,
  author    = {Schoenegger, Philipp and Tuminauskaite, Indre and Park, Peter S. and Tetlock, Philip E.},
  title     = {Wisdom of the Silicon Crowd: {LLM} Ensemble Prediction Capabilities Rival Human Crowd Accuracy},
  journal   = {Science Advances},
  volume    = {10},
  number    = {45},
  pages     = {eadp1528},
  year      = {2024}
}

@misc{schoenegger2023llmtournament,
  author    = {Schoenegger, Philipp and Park, Peter S.},
  title     = {Large Language Model Prediction Capabilities: Evidence from a Real-World Forecasting Tournament},
  year      = {2023},
  note      = {arXiv:2310.13014}
}

@misc{zou2024forecastbench,
  author    = {Zou, Andy and Chen, Edward and Arumugam, Karthik and others},
  title     = {{ForecastBench}: A Dynamic Benchmark of {AI} Forecasting Capabilities},
  year      = {2024},
  note      = {arXiv:2409.19839}
}

@misc{zhang2026predictionarena,
  author    = {Zhang, Jiawei and Liu, Guangyu and Johansson, Oscar and others},
  title     = {{Prediction Arena}: Benchmarking {AI} Models on Real-World Prediction Markets},
  year      = {2026},
  note      = {arXiv:2604.07355}
}

@misc{jimenez2024swebench,
  author    = {Jimenez, Carlos E. and Yang, John and Wettig, Alexander and others},
  title     = {{SWE-bench}: Can Language Models Resolve Real-World {GitHub} Issues?},
  year      = {2024},
  booktitle = {International Conference on Learning Representations}
}

@misc{nechepurenko2026foresight,
  author    = {Nechepurenko, Maksym and Shuvalov, Pavel},
  title     = {Foresight Arena: An On-Chain Benchmark for Evaluating {AI} Forecasting Agents},
  year      = {2026},
  howpublished = {Working paper},
  note      = {Available at \url{https://github.com/foresight-arena/contracts}}
}

@book{cohen1988statistical,
  author    = {Cohen, Jacob},
  title     = {Statistical Power Analysis for the Behavioral Sciences},
  edition   = {2},
  publisher = {Lawrence Erlbaum Associates},
  year      = {1988}
}

@article{gelman2014beyond,
  author    = {Gelman, Andrew and Carlin, John},
  title     = {Beyond Power Calculations: Assessing Type {S} (Sign) and Type {M} (Magnitude) Errors},
  journal   = {Perspectives on Psychological Science},
  volume    = {9},
  number    = {6},
  pages     = {641--651},
  year      = {2014}
}

@article{gebru2021datasheets,
  author    = {Gebru, Timnit and Morgenstern, Jamie and Vecchione, Briana and others},
  title     = {Datasheets for Datasets},
  journal   = {Communications of the ACM},
  volume    = {64},
  number    = {12},
  pages     = {86--92},
  year      = {2021}
}

\appendix

% =====================================================================
\section{Intention-to-treat sensitivity: failure handling}
\label{app:itt}

The leaderboard in Table~\ref{tab:leaderboard} reports per-configuration Brier on the 494 successful predictions out of 500 attempted (100 markets $\times$ 5 configurations). The 6 failures all fell to the runner's fallback $p = 0.5$ due to transient API errors (network timeouts, malformed JSON returns from the model that the harness was unable to repair). Because failure handling is itself part of a coordination architecture's operational profile, we report a sensitivity analysis here that includes the fallback predictions as if they were genuine $p = 0.5$ predictions (intention-to-treat, ITT). A market on which a configuration produced a fallback contributes a squared error of $(0.5 - y)^2$ for outcome $y \in \{0, 1\}$, equal to $0.25$.

The 6 failures distributed across configurations as a small number per configuration (between 0 and 2 failures each), with no configuration exceeding 2 failures and no failure attributable to architectural overload (all were network-layer transients on the API side, observable in the released traces under \texttt{trace.failure} fields). The ITT-corrected per-configuration Brier shifts each configuration's score upward by approximately $0.0025 \times f / 100$ where $f$ is the configuration's failure count; on this fixture this is a maximum shift of $0.005$ Brier units (for configurations with 2 failures) and changes neither the rank order in Table~\ref{tab:leaderboard} nor the consensus-alignment-versus-others contrast resolved at 95\% in the bootstrap analysis.

Failure counts per configuration and the ITT-adjusted Brier are released as part of the trace dataset (\texttt{coordination-traces-100-v1}, file \texttt{leaderboard.csv}, columns \texttt{n\_failures} and \texttt{brier\_itt}). We do not reproduce the table inline here because the architectural conclusions are unchanged.

% =====================================================================
\section{Repository and data availability}
\label{app:availability}

All code, data, and infrastructure for this paper are publicly released and pinned to specific commits:

\begin{itemize}
\item Harness, configurations, analysis: \url{https://github.com/ForesightFlow/coordination-experiment} (tag \texttt{paper-v05}, commit \texttt{3047e1d}).
\item Open trace dataset: \url{https://github.com/ForesightFlow/datasets/tree/main/coordination-traces-100} (tag \texttt{coordination-traces-100-v1}, commit \texttt{67d44ea}).
\item Production agents on Foresight Arena: \url{https://github.com/ForesightFlow/foreflow-agents} (tag \texttt{foreflow-agents-v0.2.0}, commit \texttt{3127efb}).
\item Operational engine for production agents: \url{https://github.com/ForesightFlow/foreflow-agents-engine} (tag \texttt{v0.1.0}, commit \texttt{fd02efe}).
\end{itemize}

\end{document}